\documentclass[12pt,letterpaper]{article}
\usepackage[table]{xcolor}
\definecolor{selectedrow}{HTML}{EDEDED}
\usepackage{float}
\usepackage{enumitem}
\PassOptionsToPackage{table}{xcolor}
\usepackage[letterpaper, total={7in, 10in}]{geometry}

\usepackage{graphicx}
\usepackage{helvet}
\usepackage{authblk}
\usepackage{hyperref}
\usepackage{amsmath} 
\usepackage{amssymb} 
\usepackage{orcidlink} 
\usepackage[super,comma,sort&compress]  
   {natbib}
\usepackage[title]{appendix}%

\usepackage[T1]{fontenc}    
\usepackage{url}            
\usepackage{booktabs}       
\usepackage{amsfonts}       
\usepackage{nicefrac}       
\usepackage{microtype}      
\usepackage[table]{xcolor}         
\usepackage{multirow}
\usepackage{threeparttable}
\usepackage{amsmath}
\usepackage{subcaption}
\usepackage{makecell}
\usepackage{placeins}
\usepackage{booktabs}      
\usepackage{longtable}
\usepackage{siunitx}       
\usepackage{caption}       
\usepackage{tabularx}
\usepackage{listings}    
\usepackage{pifont}
\usepackage{array}
\usepackage{makecell}
\theadset{\bfseries}

\makeatletter
\renewcommand{\maketitle}{\bgroup\setlength{\parindent}{0pt}
\begin{flushleft}
  \textbf{\@title}
  
  \@author
\end{flushleft}\egroup}
\makeatother

\title{ECG-LabWAS: A Laboratory-Wide Association Study Using a Fine-Tuned Single-Lead Electrocardiogram Foundation Model}
\date{}

\author[1,2,\#]{Wenhui Liu}
\author[1,2,3,\#]{Yujie Xiao}
\author[1,2,\#]{Gongzheng Tang}
\author[2]{Jianwei Chen}
\author[1,2]{Jun Li}
\author[1,2,3]{Guangkun Nie}
\author[4]{Deyun Zhang}
\author[5,*]{Qinghao Zhao}
\author[1,2,6,7,*,\orcidlink{0000-0001-7521-5127}]{Shenda Hong}

\affil[1]{Institute of Medical Technology, Peking University Health Science Center, Beijing, China}
\affil[2]{National Institute of Health Data Science, Peking University, Beijing, China}
\affil[3]{School of Intelligence Science and Technology, Peking University, Beijing, China}
\affil[4]{HeartVoice Medical Technology, Hefei, China}
\affil[5]{Department of Cardiology, Peking University People's Hospital, Beijing, China}
\affil[6]{Institute for Artificial Intelligence, Peking University, Beijing, China}
\affil[7]{State Key Laboratory of Vascular Homeostasis and Remodeling, NHC Key Laboratory of Cardiovascular Molecular Biology and Regulatory Peptides, Peking University, Beijing, China}

\affil[ ]{\textsuperscript{\#}These authors contributed equally to this work.}
\affil[ ]{\textsuperscript{*}Correspondence: hongshenda@pku.edu.cn, qhzhao@pku.edu.cn}

\begin{document}

\maketitle

\section*{ABSTRACT}

Laboratory testing guides clinical diagnosis but requires blood
sampling and processing. The electrocardiogram (ECG) is fast and
noninvasive and may carry information about laboratory
abnormalities, yet most AI-ECG models target one test at a time
using 12-lead recordings. We conducted an exploratory ECG
laboratory-wide association study (ECG-LabWAS), fine-tuning an ECG
foundation model pretrained on 10.8 million ECGs to screen 50
laboratory tests from single-lead waveforms alone. The model was
developed in an emergency-department cohort (MC-MED; 31,826
patients) and applied unchanged to an independent hospital cohort
(MIMIC-IV; 115,921 patients). Of 89 abnormality directions (low or
high values of each test), 33 met the same screening criteria in
both cohorts. Areas under the receiver operating characteristic
curve in the two cohorts were 0.841 and 0.838 for elevated
NT-proBNP, 0.769 and 0.740 for elevated cardiac troponin, 0.755 and
0.731 for elevated creatinine, and 0.713 and 0.710 for low
potassium. Discrimination changed little when the ECG--laboratory
matching window was widened from $\pm$1 to $\pm$24 hours. Model
predictions reproduced the co-occurrence of observed abnormalities
but were about twice as tightly coupled, largely along one shared
component consistent with overall illness severity; low potassium,
elevated NT-proBNP and troponin, and markers of renal function
retained discrimination beyond that of correlated tests. In
exploratory analyses, ECG-assisted versions of the modified Glasgow
Prognostic Score and the systemic inflammatory response syndrome
criteria were associated with in-hospital and one-year mortality.
These results identify which laboratory abnormalities warrant
analyte-specific study with single-lead ECG.

\section*{SIGNIFICANCE}

Blood tests guide many medical decisions, but they require drawing
blood and waiting for laboratory processing. A single-lead
electrocardiogram, the heart tracing recorded by bedside monitors
and many wearable devices, is available within seconds. We asked
which laboratory abnormalities leave a detectable trace in this
simple signal. An artificial intelligence model trained in one
health system and applied unchanged in another detected 33
abnormal laboratory results in both, including markers of heart
strain, kidney function, and low potassium. Much of the signal
reflected how ill a patient was overall rather than any single
test, but several abnormalities carried more specific information.
These findings identify the most promising targets for ECG-based
laboratory monitoring.

\section*{KEYWORDS}

Electrocardiography; deep learning; single-lead ECG; clinical
laboratory tests.


\section*{INTRODUCTION}

Many diagnostic and treatment decisions depend on laboratory test
results\cite{aghaei2024need,gao2024laboratory}. These results
usually require a venous blood sample processed in a central
laboratory\cite{lima2015laboratory}. Sampling is
invasive\cite{whitehead2019interventions}, and collection,
transport, and assay processing each add
delay\cite{vrijsen2022shorter}. The cost is greatest when a decision cannot wait or when repeated sampling is undesirable. In acute care, an electrocardiogram (ECG) may already be on hand while laboratory results are still pending.

The ECG is widely available, inexpensive, and non-invasive, and it
takes only minutes to record\cite{rafie2021ecg}. It measures cardiac electrical activity, but the waveform may also reflect systemic changes that accompany abnormal laboratory test results (hereafter, laboratory abnormalities). Deep learning has advanced the analysis of medical images and physiological signals\cite{chen2024towards,li2025expert}, and AI-ECG models are
now used for disease detection and risk stratification\cite{mayourian2025electrocardiogram, mayourian2025electrocardiogram1,khurshid2022ecg,
strodthoff2020deep,jiang2024deep,xiao2026fine}. For laboratory tests specifically, dedicated limited-lead models can detect abnormal potassium from one or two leads\cite{galloway2019development,chiu2024serum}, and CardioLab and related work have evaluated broader panels that include NT-proBNP\cite{alcaraz2025abnormality}. ECG foundation models such as ECGFounder, pretrained on large ECG collections, provide representations that can be fine-tuned for downstream tasks\cite{li2025electrocardiogram}. Even so, most studies
have examined single analytes or selected panels. Although a single lead is what bedside monitors and wearables provide, whether a broad laboratory panel can be detected from it across different health systems has not been examined systematically.

Two gaps limit what existing work can tell us about single-lead
ECGs. First, most models were developed on 12-lead recordings and
target one endpoint, and their performance does not transfer
automatically to a single lead\cite{avula2023clinical,presacan2025evaluating}.
Second, few models have been tested in an independent health
system\cite{avula2023clinical,alcaraz2025abnormality}, so a signal
found in one hospital may reflect local patients or testing
practices. As a result, it is unclear which laboratory abnormalities can be detected from a single lead in more than one institution. Screening a broad panel with one model, and then applying that model unchanged in a second health system, would show which candidates are worth studying further.

We developed ECG-LabWAS to systematically identify laboratory
abnormalities associated with single-lead ECGs, drawing on the
panel-wide screening approach of phenome-wide association studies.
We fine-tuned ECGFounder with ten classification heads, grouped by
the presumed pathway linking each abnormality to the ECG, on the Multimodal Clinical Monitoring in the Emergency Department dataset (MC-MED)\cite{PhysioNet-mc-med-1.0.1,kansal2025mc} and evaluated the selected checkpoint in MIMIC-IV\cite{johnson2023mimic,gow2023mimic} without retraining or external model selection. The model received ECG waveforms alone and covered a prespecified panel of 50 laboratory tests comprising 89 abnormality directions. Applying the same patient-first screening criteria in both cohorts identified 33 directions for further investigation. We then examined the stability of discrimination across training seeds and ECG--laboratory matching windows. To ask whether the model detected each abnormality specifically or through a signal shared across abnormalities, we compared the coupling among the model's predictions with the coupling among the observed laboratory labels. Finally, we explored the prognostic relevance of the model outputs through an ECG-derived modified Glasgow Prognostic Score
(ECG-mGPS), based on the mGPS\cite{forrest2003evaluation,mcmillan2007evaluation},
and a hybrid ECG-assisted score based on the systemic inflammatory
response syndrome (SIRS) criteria\cite{bone1992definitions} (hybrid ECG-SIRS), assessing their associations with in-hospital and one-year mortality in MIMIC-IV.


\section*{RESULTS}

\subsection*{Fine-tuned ECGFounder enables laboratory-wide
screening from single-lead ECGs}
\label{sec:framework}
\begin{figure}[htbp]
\centering
\includegraphics[width=\linewidth]
{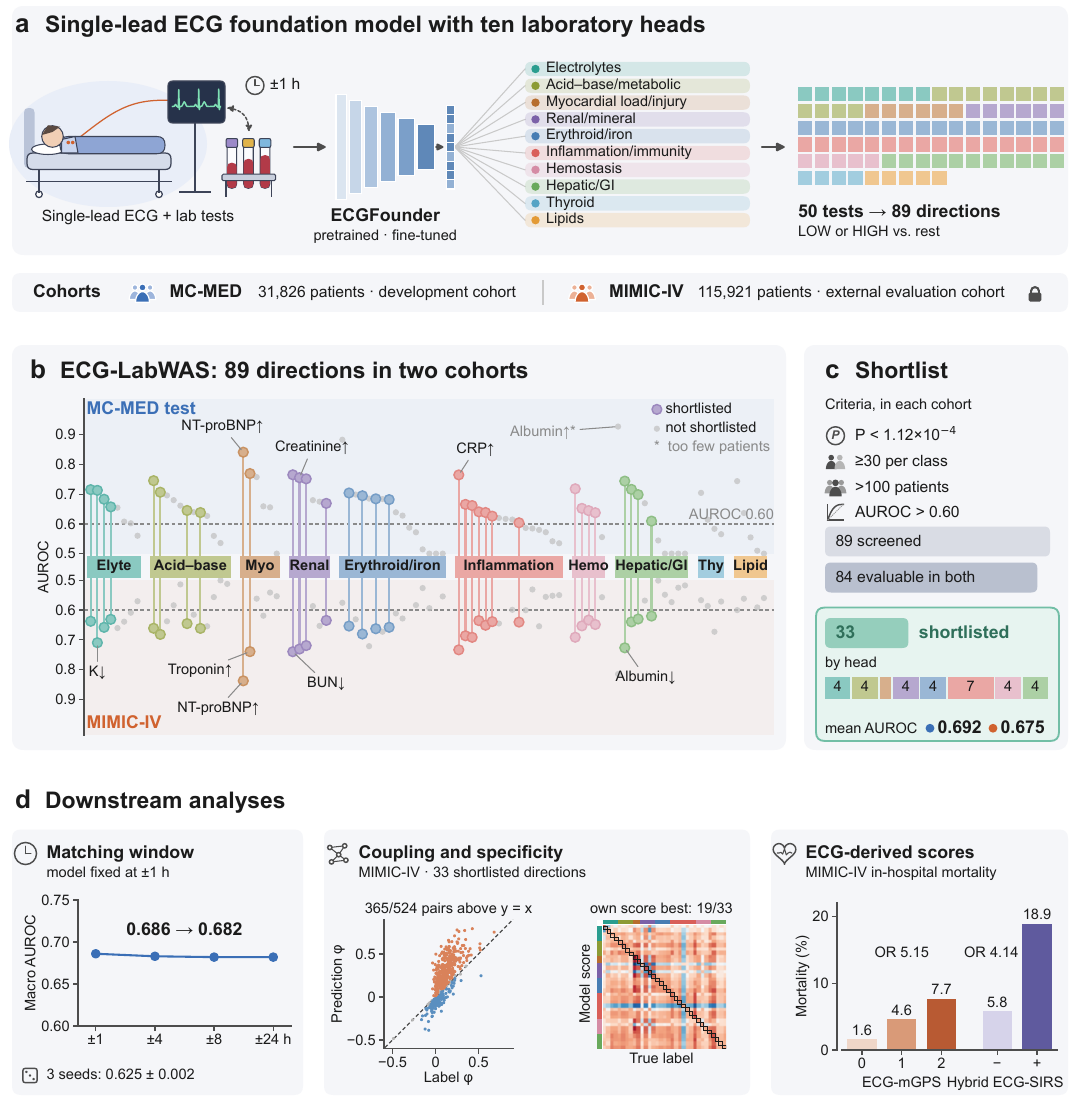}
\caption{\textbf{ECG-LabWAS overview.}
\textbf{(a)} ECGFounder was fine-tuned on single-lead ECGs paired with
laboratory results within $\pm$1 hour in MC-MED, with ten heads
covering 50 tests and 89 abnormality directions. The selected
checkpoint was applied unchanged to MIMIC-IV.
\textbf{(b)} Patient-first AUROC for each direction in MC-MED (top) and
MIMIC-IV (bottom). Colored, linked points are the 33 shortlisted
directions; gray points are not. Values below 0.5 are drawn at the
baseline. *, too few patients. Elyte, electrolytes; Myo, myocardial
load and injury; Hemo, hemostasis; Thy, thyroid; GI, gastrointestinal.
\textbf{(c)} Shortlist criteria, applied in each cohort.
\textbf{(d)} Matching-window sensitivity in MC-MED (seed value: mean
task-level AUROC in MIMIC-IV across three seeds), coupling and
specificity of the 33 shortlisted directions in MIMIC-IV (left,
observed-label versus prediction $\varphi$ for each pair, with the
dashed line $y=x$; right, cross-direction AUROC with model scores in
rows and true labels in columns, outlined cells marking each
direction's own score; see Fig.~\ref{fig:coupling_specificity}), and in-hospital
mortality in MIMIC-IV by ECG-mGPS and hybrid ECG-SIRS, with age- and
sex-adjusted odds ratios.}
\label{fig:architecture}
\end{figure}

We established a laboratory-wide screening framework by fine-tuning single-lead ECGFounder on MC-MED and evaluating it in MIMIC-IV (Fig.~\ref{fig:architecture}a--c). Ten classification heads,
grouped by the presumed pathway linking each abnormality to the ECG,
covered 50 laboratory tests and 89 one-versus-rest abnormality directions. Each model sample paired
an ECG waveform with laboratory results obtained within
$\pm$1 hour; the model received only the waveform. The checkpoint
selected on the MC-MED validation partition was evaluated in the
held-out MC-MED test partition and applied unchanged to single-lead waveforms extracted from MIMIC-IV diagnostic ECGs. Using the
earliest observed record for each patient and task, we applied the
same statistical, sample-size, and discrimination criteria in both
cohorts to define a cross-cohort shortlist. Subsequent analyses
examined the stability of discrimination, the coupling between
observed and predicted abnormalities, and the mortality
associations of ECG-based clinical scores
(Fig.~\ref{fig:architecture}d).

The ECG-matched MC-MED cohort included 31{,}826 patients and
115{,}405 model samples (Table~\ref{tab:population}). The training,
validation, and test partitions contained 22{,}275, 6{,}364, and
3{,}187 patients, with 80{,}590, 23{,}586, and 11{,}229 model samples
and 65{,}220, 19{,}172, and 9{,}066 unique ECG inputs, respectively.
Mean age was 57.9 years, and 51.8\% of patients were female. The
MIMIC-IV cohort included 300{,}035 unique ECGs from 115{,}921
patients across 47 mapped laboratory tasks; mean age was 59.9 years,
and 50.9\% of patients were female. The cohorts differed in demographic composition and in how ECGs were acquired, with MC-MED providing continuous bedside-monitor recordings and MIMIC-IV providing 12-lead diagnostic ECGs.
Coded comorbidities were broadly similar, although dyslipidemia, chronic kidney disease,
obesity, and thyroid disease were recorded less often in MIMIC-IV (Table~\ref{tab:population}).

\begin{figure}[htbp]
\centering
\includegraphics[width=\linewidth]
{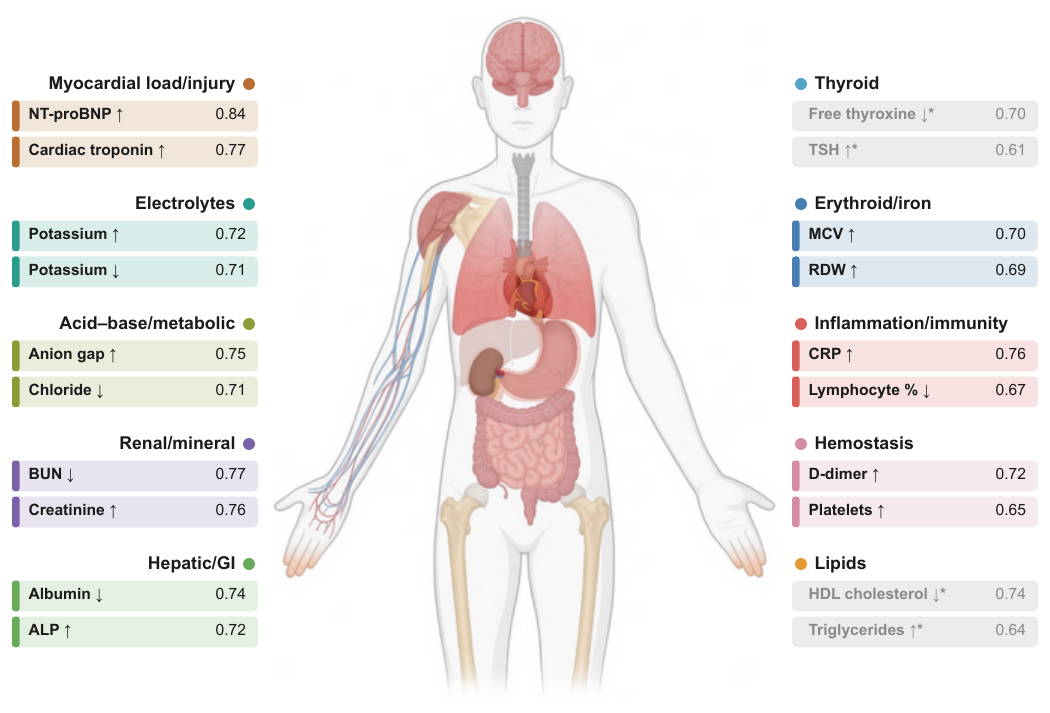}
\caption{\textbf{Representative laboratory abnormality directions by classification head.} Each row shows one abnormality direction from each of the ten heads ($\uparrow$, high; $\downarrow$, low) and its patient-first AUROC in the MC-MED test partition. Heads group tests by the presumed pathway linking an abnormality to the ECG; the anatomical illustration is decorative and does not map heads to organs. Gray rows marked with an asterisk did not meet the shortlist criteria; the thyroid and lipid heads had no shortlisted direction, and their AUROCs rest on few test patients. The 33 shortlisted directions are listed in Table~\ref{tab:cross_cohort}, and the full panel with 95\% CIs in Supplementary Table~\ref{tab:s1}.}
\label{fig:anatomy}
\end{figure}

Figure~\ref{fig:architecture}b,c summarizes the screen, and Figure~\ref{fig:anatomy} shows representative directions for each head, including directions that did not meet the shortlist criteria.

\begin{table}[t]
\centering
\caption{\textbf{Patient-level characteristics of the ECG-LabWAS
analysis cohorts.}
Values are $n$ (\%). Train, Valid, Test, and Overall refer to
MC-MED; MIMIC-IV is the external cohort. Age is categorized as
$>60$ versus $\le60$ years at the earliest model input.
Mean age was 57.9 years in MC-MED and 59.9 years in MIMIC-IV.
Race and ethnicity were standardized across cohorts, with
Hispanic or Latino ethnicity taking precedence in MC-MED.
For MIMIC-IV, race and ethnicity were taken as the most frequent
known category across hospital admissions and emergency-department
stays. Comorbidities are coded diagnoses recorded at or before each
patient's earliest model ECG (MC-MED past medical history; MIMIC-IV
hospital and emergency-department diagnoses). Patients without a
matching code are counted as not having that condition. Only the
MC-MED training partition was used to update model parameters.}
\label{tab:population}
\footnotesize
\setlength{\tabcolsep}{3pt}
\renewcommand{\arraystretch}{1.05}
\rowcolors{2}{black!5}{white}
\resizebox{\linewidth}{!}{%
\begin{tabular}{@{}lccccc@{}}
\toprule
Characteristic & Train & Valid & Test & Overall & MIMIC-IV \\
\midrule
Patients, $n$ &
22{,}275 & 6{,}364 & 3{,}187 & 31{,}826 & 115{,}921 \\
\addlinespace
Age, $n$ (\%) & & & & & \\
\quad $>60$ years &
10{,}823 (48.6) & 3{,}083 (48.4) & 1{,}545 (48.5) &
15{,}451 (48.5) & 60{,}049 (51.8) \\
\quad $\le60$ years &
11{,}452 (51.4) & 3{,}281 (51.6) & 1{,}642 (51.5) &
16{,}375 (51.5) & 55{,}872 (48.2) \\
\addlinespace
Sex, $n$ (\%) & & & & & \\
\quad Female &
11{,}525 (51.7) & 3{,}268 (51.4) & 1{,}688 (53.0) &
16{,}481 (51.8) & 59{,}040 (50.9) \\
\quad Male &
10{,}738 (48.2) & 3{,}090 (48.6) & 1{,}498 (47.0) &
15{,}326 (48.2) & 56{,}881 (49.1) \\
\quad Other / Missing &
12 (0.1) & 6 (0.1) & 1 (0.0) & 19 (0.1) & 0 (0.0) \\
\addlinespace
Race/ethnicity, $n$ (\%) & & & & & \\
\quad White &
9{,}228 (41.4) & 2{,}599 (40.8) & 1{,}266 (39.7) &
13{,}093 (41.1) & 75{,}771 (65.4) \\
\quad Black / African American &
1{,}147 (5.1) & 343 (5.4) & 158 (5.0) &
1{,}648 (5.2) & 15{,}884 (13.7) \\
\quad Asian &
4{,}049 (18.2) & 1{,}120 (17.6) & 580 (18.2) &
5{,}749 (18.1) & 4{,}055 (3.5) \\
\quad Hispanic or Latino &
5{,}114 (23.0) & 1{,}490 (23.4) & 759 (23.8) &
7{,}363 (23.1) & 6{,}185 (5.3) \\
\quad Other &
2{,}700 (12.1) & 800 (12.6) & 423 (13.3) &
3{,}923 (12.3) & 4{,}986 (4.3) \\
\quad Unknown / Missing &
37 (0.2) & 12 (0.2) & 1 (0.0) &
50 (0.2) & 9{,}040 (7.8) \\
\addlinespace
Comorbidity, $n$ (\%) & & & & & \\
\quad Hypertension &
4{,}908 (22.0) & 1{,}449 (22.8) & 714 (22.4) &
7{,}071 (22.2) & 26{,}910 (23.2) \\
\quad Diabetes mellitus &
2{,}407 (10.8) & 732 (11.5) & 362 (11.4) &
3{,}501 (11.0) & 11{,}855 (10.2) \\
\quad Dyslipidemia &
3{,}916 (17.6) & 1{,}164 (18.3) & 563 (17.7) &
5{,}643 (17.7) & 16{,}067 (13.9) \\
\quad Coronary artery disease &
2{,}130 (9.6) & 574 (9.0) & 292 (9.2) &
2{,}996 (9.4) & 11{,}725 (10.1) \\
\quad Heart failure &
1{,}493 (6.7) & 432 (6.8) & 194 (6.1) &
2{,}119 (6.7) & 6{,}093 (5.3) \\
\quad Atrial fibrillation or flutter &
1{,}870 (8.4) & 504 (7.9) & 248 (7.8) &
2{,}622 (8.2) & 7{,}967 (6.9) \\
\quad Chronic kidney disease &
1{,}528 (6.9) & 429 (6.7) & 208 (6.5) &
2{,}165 (6.8) & 5{,}239 (4.5) \\
\quad Chronic liver disease &
1{,}080 (4.8) & 312 (4.9) & 144 (4.5) &
1{,}536 (4.8) & 3{,}406 (2.9) \\
\quad Chronic pulmonary disease &
744 (3.3) & 190 (3.0) & 96 (3.0) &
1{,}030 (3.2) & 3{,}905 (3.4) \\
\quad Obesity &
1{,}326 (6.0) & 402 (6.3) & 227 (7.1) &
1{,}955 (6.1) & 4{,}015 (3.5) \\
\quad Anemia &
2{,}713 (12.2) & 743 (11.7) & 389 (12.2) &
3{,}845 (12.1) & 11{,}916 (10.3) \\
\quad Thyroid disease &
1{,}877 (8.4) & 525 (8.2) & 272 (8.5) &
2{,}674 (8.4) & 5{,}288 (4.6) \\
\bottomrule
\end{tabular}%
}
\end{table}

\subsection*{Cross-cohort screening identifies 33 laboratory
abnormality directions}
\label{sec:shortlist}

On the held-out MC-MED test partition, 88 of the 89 abnormality
directions could be evaluated, with a mean patient-first area under
the receiver operating characteristic curve (AUROC) of 0.614. Supplementary Table~\ref{tab:s1} lists every direction,
including those with few observations, low discrimination, or no
external mapping.

Discrimination varied widely. Among the 53 directions with at least
30 positive patients, at least 30 negative patients, and more than
100 observed patients, 14 had an AUROC above 0.70: elevated NT-proBNP
(0.841, 95\% CI 0.807--0.874), elevated cardiac troponin
(0.769, 95\% CI 0.733--0.803), low blood urea nitrogen
(BUN; 0.765, 95\% CI 0.699--0.827), elevated C-reactive protein
(CRP; 0.765, 95\% CI 0.687--0.838), elevated creatinine
(0.755, 95\% CI 0.732--0.780), elevated BUN
(0.752, 95\% CI 0.731--0.772), elevated anion gap
(0.745, 95\% CI 0.706--0.782), low albumin
(0.744, 95\% CI 0.718--0.771), elevated D-dimer
(0.718, 95\% CI 0.666--0.771), elevated alkaline phosphatase
(0.716, 95\% CI 0.690--0.743), elevated potassium
(0.715, 95\% CI 0.676--0.753), low potassium
(0.713, 95\% CI 0.676--0.747), low chloride
(0.706, 95\% CI 0.675--0.735), and elevated mean corpuscular volume
(0.704, 95\% CI 0.668--0.736).

AUROC values from sparsely observed directions are unreliable.
Elevated albumin reached 0.927 (95\% CI 0.917--0.937), but this value rested on very few
positive patients and did not survive multiplicity correction.
Across evaluable directions, AUROC was weakly correlated
with the number of positive training samples (Spearman $\rho=0.28$,
$P=0.008$; Supplementary Fig.~\ref{fig:s1}), and sample size
still affects the precision of each estimate.

Mean AUROC was 0.655 among
the 53 directions that met the sample-size criteria and
0.683 among
the 38 directions that met the Bonferroni criterion; because these
subsets were defined by different filters, the differences between
these means reflect the filters rather than better discrimination. To enter the shortlist, a
direction also needed an AUROC above 0.60 and had to meet every
criterion in MIMIC-IV.

In MIMIC-IV, the frozen checkpoint could be evaluated for 85 of the
89 directions, and 74 met the Bonferroni criterion (family-wise
$\alpha=0.01$). Eighty-four directions were evaluable in both
cohorts, and 33 met all statistical, sample-size, and discrimination criteria independently in both
cohorts (Table~\ref{tab:cross_cohort}): 7 from the inflammation and
immunity head; 4 each from the electrolyte, acid--base and metabolic,
renal and mineral, erythroid and iron, hemostasis, and hepatic and
gastrointestinal heads; and 2 from the myocardial head (elevated
NT-proBNP and elevated cardiac troponin).
Their mean AUROC was 0.692 in MC-MED
(range across directions, 0.604--0.841) and
0.675 in MIMIC-IV
(range across directions, 0.620--0.838).
In MIMIC-IV, the troponin label combined troponin I and troponin T
results (Methods); elevated cardiac troponin had an AUROC of 0.740
(95\% CI 0.735--0.744) among 58{,}101 patients.

Performance changed between cohorts in different ways for different
directions. Elevated NT-proBNP had the highest external AUROC
(0.838, 95\% CI 0.830--0.847). Low BUN
(0.740, 95\% CI 0.725--0.754), low albumin
(0.727, 95\% CI 0.721--0.733), and elevated BUN
(0.719, 95\% CI 0.716--0.722) retained similar point estimates
across cohorts. The AUROC for elevated white blood cell count
was 0.662 (95\% CI 0.634--0.687) in MC-MED and
0.692 (95\% CI 0.687--0.696) in MIMIC-IV. Corresponding
values were 0.745 (95\% CI 0.706--0.782) and
0.661 (95\% CI 0.657--0.665) for elevated anion gap,
0.715 (95\% CI 0.676--0.753) and
0.637 (95\% CI 0.631--0.644) for elevated potassium, and
0.716 (95\% CI 0.690--0.743) and
0.640 (95\% CI 0.632--0.646) for elevated alkaline phosphatase.
These are descriptive comparisons of cohort-specific estimates.
Given this heterogeneity, we report estimates for each direction
rather than relying on the shortlist average.

\clearpage
\begingroup
\scriptsize
\rowcolors{1}{black!5}{white}
\setlength{\tabcolsep}{1.8pt}
\begin{longtable}{
@{}
>{\raggedright\arraybackslash}p{0.22\linewidth}
l
>{\raggedright\arraybackslash}p{0.19\linewidth}
c r c r
@{}
}
\rowcolor{white} 
\caption{\textbf{Shortlist of 33 ECG-LabWAS directions.}
Patient-first AUROC with stratified bootstrap 95\% confidence
intervals. Each $n$ is the number of evaluable patients, and the
percentage is the positive rate among them. Rates below 0.1\% are
shown as $<$0.1\%. Every direction met the
same Bonferroni significance, class-count, total-observation, and
AUROC criteria in both cohorts. Because both cohorts were used for
selection, the estimates are descriptive. The full panel appears in
Supplementary Table~\ref{tab:s1}. \textsuperscript{a}Troponin I in MC-MED; troponin I or troponin T in MIMIC-IV, each judged against its own threshold (see Methods).}
\label{tab:cross_cohort}\\
\toprule
Laboratory test & Dir. & Head &
\multicolumn{2}{c}{MC-MED} &
\multicolumn{2}{c}{MIMIC-IV} \\
\cmidrule(lr){4-5}\cmidrule(lr){6-7}
& & & AUROC (95\% CI) & $n$ (\%) & AUROC (95\% CI) & $n$ (\%) \\
\midrule
\endfirsthead

\rowcolor{white} 
\caption[]{\textbf{Shortlist of 33 ECG-LabWAS directions (continued).}}\\
\toprule
Laboratory test & Dir. & Head &
\multicolumn{2}{c}{MC-MED} &
\multicolumn{2}{c}{MIMIC-IV} \\
\cmidrule(lr){4-5}\cmidrule(lr){6-7}
& & & AUROC (95\% CI) & $n$ (\%) & AUROC (95\% CI) & $n$ (\%) \\
\midrule
\endhead

\midrule
\multicolumn{7}{r}{\emph{Continued on next page}}\\
\endfoot

\bottomrule
\endlastfoot

NT-proBNP & HIGH & Myocardial & 0.841 (0.807--0.874) & 602 (75.4\%) & 0.838 (0.830--0.847) & 13{,}621 (84.6\%) \\
Cardiac troponin\textsuperscript{a} & HIGH & Myocardial & 0.769 (0.733--0.803) & 1{,}516 (11.1\%) & 0.740 (0.735--0.744) & 58{,}101 (23.4\%) \\
Blood urea nitrogen (BUN) & LOW & Renal/mineral & 0.765 (0.699--0.827) & 2{,}856 (1.4\%) & 0.740 (0.725--0.754) & 105{,}783 (1.0\%) \\
C-reactive protein (CRP) & HIGH & Inflammation/immunity & 0.765 (0.687--0.838) & 179 (63.7\%) & 0.734 (0.715--0.752) & 3{,}150 (50.8\%) \\
Creatinine & HIGH & Renal/mineral & 0.755 (0.732--0.780) & 2{,}848 (13.8\%) & 0.731 (0.727--0.735) & 105{,}986 (15.8\%) \\
Blood urea nitrogen (BUN) & HIGH & Renal/mineral & 0.752 (0.731--0.772) & 2{,}856 (24.9\%) & 0.719 (0.716--0.722) & 105{,}783 (32.3\%) \\
Anion gap & HIGH & Acid--base/metabolic & 0.745 (0.706--0.782) & 2{,}848 (6.6\%) & 0.661 (0.657--0.665) & 102{,}396 (29.2\%) \\
Albumin & LOW & Hepatic/GI & 0.744 (0.718--0.771) & 2{,}822 (12.3\%) & 0.727 (0.721--0.733) & 37{,}641 (20.0\%) \\
D-dimer & HIGH & Hemostasis & 0.718 (0.666--0.771) & 371 (54.7\%) & 0.691 (0.680--0.702) & 8{,}282 (38.9\%) \\
Alkaline phosphatase & HIGH & Hepatic/GI & 0.716 (0.690--0.743) & 2{,}821 (13.3\%) & 0.640 (0.632--0.646) & 46{,}318 (15.9\%) \\
Potassium & HIGH & Electrolytes & 0.715 (0.676--0.753) & 2{,}851 (6.4\%) & 0.637 (0.631--0.644) & 103{,}377 (8.9\%) \\
Potassium & LOW & Electrolytes & 0.713 (0.676--0.747) & 2{,}851 (8.6\%) & 0.710 (0.704--0.715) & 103{,}377 (7.1\%) \\
Chloride & LOW & Acid--base/metabolic & 0.706 (0.675--0.735) & 2{,}848 (10.5\%) & 0.682 (0.676--0.688) & 103{,}625 (7.4\%) \\
Mean corpuscular volume (MCV) & HIGH & Erythroid/iron & 0.704 (0.668--0.736) & 2{,}986 (5.9\%) & 0.654 (0.646--0.661) & 106{,}071 (4.5\%) \\
Globulin & HIGH & Hepatic/GI & 0.698 (0.673--0.725) & 2{,}820 (15.6\%) & 0.630 (0.597--0.661) & 4{,}635 (6.9\%) \\
Red cell distribution width (RDW) & HIGH & Erythroid/iron & 0.695 (0.672--0.715) & 2{,}985 (26.5\%) & 0.680 (0.677--0.684) & 106{,}061 (25.9\%) \\
Hematocrit (HCT) & LOW & Erythroid/iron & 0.684 (0.662--0.705) & 2{,}993 (26.7\%) & 0.663 (0.659--0.666) & 106{,}809 (31.5\%) \\
Calcium & LOW & Electrolytes & 0.683 (0.652--0.717) & 2{,}849 (8.3\%) & 0.658 (0.652--0.664) & 52{,}896 (20.2\%) \\
Red blood cells (RBC) & LOW & Erythroid/iron & 0.682 (0.663--0.701) & 2{,}987 (39.3\%) & 0.658 (0.655--0.661) & 106{,}071 (41.7\%) \\
Creatinine & LOW & Renal/mineral & 0.669 (0.642--0.696) & 2{,}848 (14.0\%) & 0.635 (0.626--0.643) & 105{,}986 (3.8\%) \\
Lymphocyte \% & LOW & Inflammation/immunity & 0.666 (0.646--0.685) & 2{,}788 (47.3\%) & 0.686 (0.683--0.690) & 88{,}956 (46.6\%) \\
White blood cells (WBC) & HIGH & Inflammation/immunity & 0.662 (0.634--0.687) & 2{,}996 (16.0\%) & 0.692 (0.687--0.696) & 106{,}132 (17.5\%) \\
Sodium & LOW & Electrolytes & 0.657 (0.633--0.683) & 2{,}856 (18.2\%) & 0.631 (0.626--0.636) & 103{,}159 (11.4\%) \\
Platelet count (PLT) & HIGH & Hemostasis & 0.652 (0.615--0.691) & 2{,}995 (7.0\%) & 0.653 (0.646--0.660) & 106{,}251 (5.8\%) \\
Glucose & HIGH & Acid--base/metabolic & 0.645 (0.622--0.667) & 2{,}856 (72.3\%) & 0.645 (0.642--0.649) & 104{,}859 (66.9\%) \\
Partial thromboplastin time & HIGH & Hemostasis & 0.645 (0.609--0.680) & 1{,}238 (23.6\%) & 0.633 (0.628--0.638) & 69{,}229 (17.9\%) \\
Immature granulocyte \% & HIGH & Inflammation/immunity & 0.641 (0.618--0.661) & 2{,}786 (36.9\%) & 0.635 (0.629--0.642) & 30{,}951 (42.8\%) \\
Eosinophil \% & LOW & Inflammation/immunity & 0.638 (0.619--0.659) & 2{,}788 (40.5\%) & 0.651 (0.647--0.655) & 88{,}947 (36.9\%) \\
Bicarbonate & LOW & Acid--base/metabolic & 0.638 (0.616--0.663) & 2{,}848 (25.6\%) & 0.661 (0.657--0.666) & 102{,}758 (21.4\%) \\
Platelet count (PLT) & LOW & Hemostasis & 0.638 (0.609--0.668) & 2{,}995 (10.3\%) & 0.648 (0.643--0.653) & 106{,}251 (10.7\%) \\
Neutrophil \% & HIGH & Inflammation/immunity & 0.626 (0.603--0.647) & 2{,}788 (70.1\%) & 0.639 (0.635--0.643) & 88{,}948 (74.6\%) \\
AST (SGOT) & HIGH & Hepatic/GI & 0.609 (0.589--0.631) & 2{,}821 (37.2\%) & 0.620 (0.615--0.625) & 47{,}905 (36.9\%) \\
Lymphocyte \% & HIGH & Inflammation/immunity & 0.604 (0.561--0.644) & 2{,}788 (6.5\%) & 0.640 (0.634--0.647) & 88{,}956 (7.6\%) \\
\end{longtable}
\endgroup

\begin{figure}[htbp]
\centering
\includegraphics[width=\linewidth]
{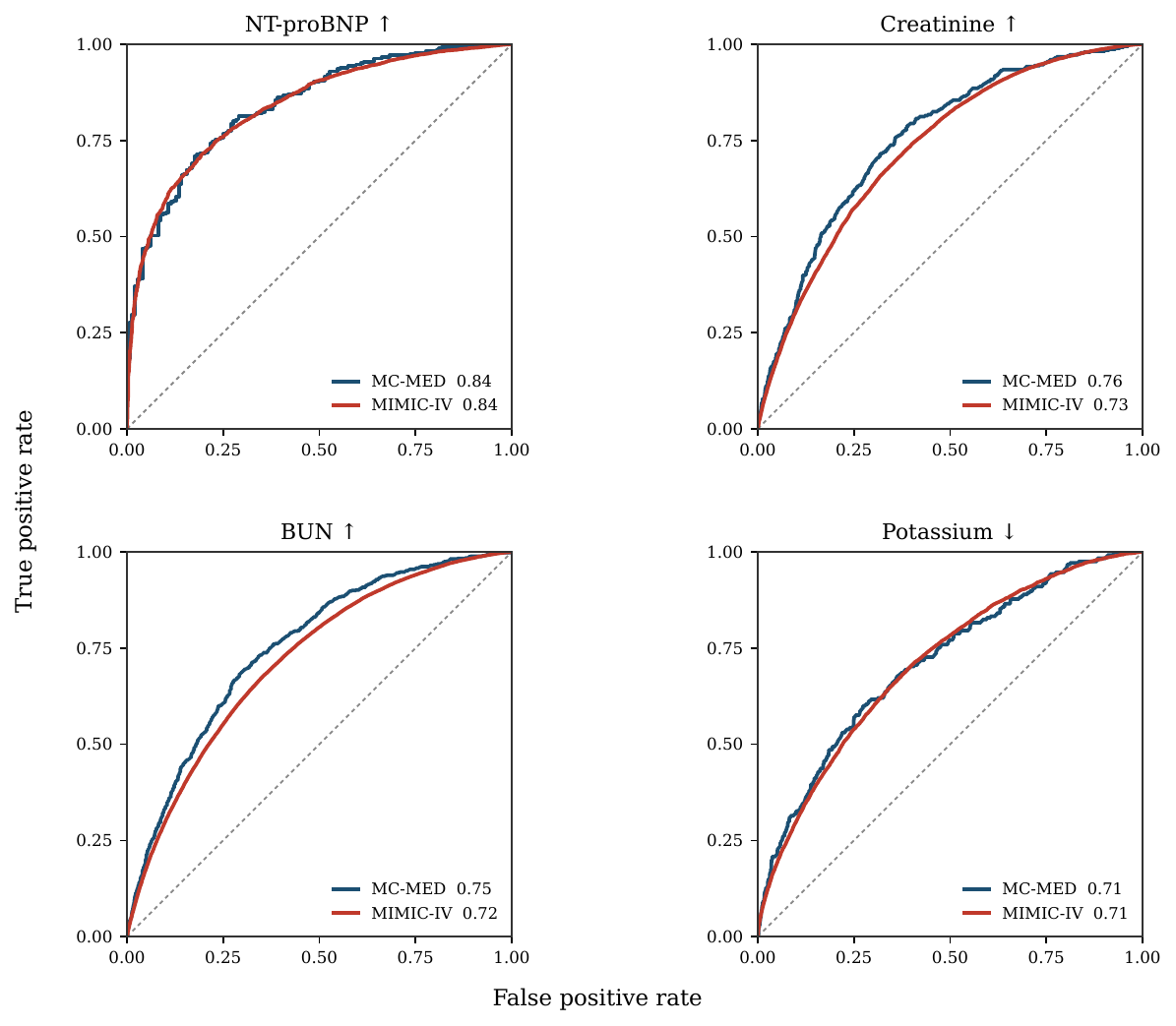}
\caption{\textbf{Patient-first receiver operating characteristic
curves for four representative directions.}
Curves compare internal testing in MC-MED with external evaluation
in MIMIC-IV for elevated NT-proBNP, elevated creatinine, elevated
blood urea nitrogen, and low potassium. The dashed diagonal marks
chance discrimination. AUROC values in the legends are rounded to two
decimal places; the corresponding three-decimal AUROCs and
95\% CIs are reported in Table~\ref{tab:cross_cohort}.}
\label{fig:roc_core}
\end{figure}

Figure~\ref{fig:roc_core} shows ROC curves for four directions with
adequate class counts and an AUROC of about 0.70 or higher in both
cohorts, chosen to cover cardiac, renal, and electrolyte abnormalities.
The MC-MED and MIMIC-IV AUROCs were, respectively,
0.841 (95\% CI 0.807--0.874) and 0.838 (95\% CI 0.830--0.847)
for elevated NT-proBNP; 0.755 (95\% CI 0.732--0.780) and
0.731 (95\% CI 0.727--0.735) for elevated creatinine;
0.752 (95\% CI 0.731--0.772) and 0.719 (95\% CI 0.716--0.722)
for elevated BUN; and 0.713 (95\% CI 0.676--0.747) and
0.710 (95\% CI 0.704--0.715) for low potassium.

\subsection*{Aggregate discrimination varies little across
training seeds and matching windows}
\label{sec:sensitivity}

After identifying the cross-cohort shortlist, we examined whether
discrimination depended on training randomness or on the temporal
alignment of ECG and laboratory measurements. We first repeated
fine-tuning with seeds 123 and 456 on the same patient partition,
using the same validation-based checkpoint selection procedure
(Supplementary Table~\ref{tab:s2}). Across seeds 42, 123, and 456,
the sample-level task-macro abnormal-class AUROC (mean $\pm$ SD across
the three seeds) was $0.615\pm0.004$ in
the MC-MED test partition and $0.625\pm0.002$ across the 47 mapped
MIMIC-IV tasks. Internal AUROCs ranged from 0.611 to 0.618, and
external AUROCs ranged from 0.623 to 0.627.

\begin{figure}[htbp]
\centering
\includegraphics[width=\linewidth]
{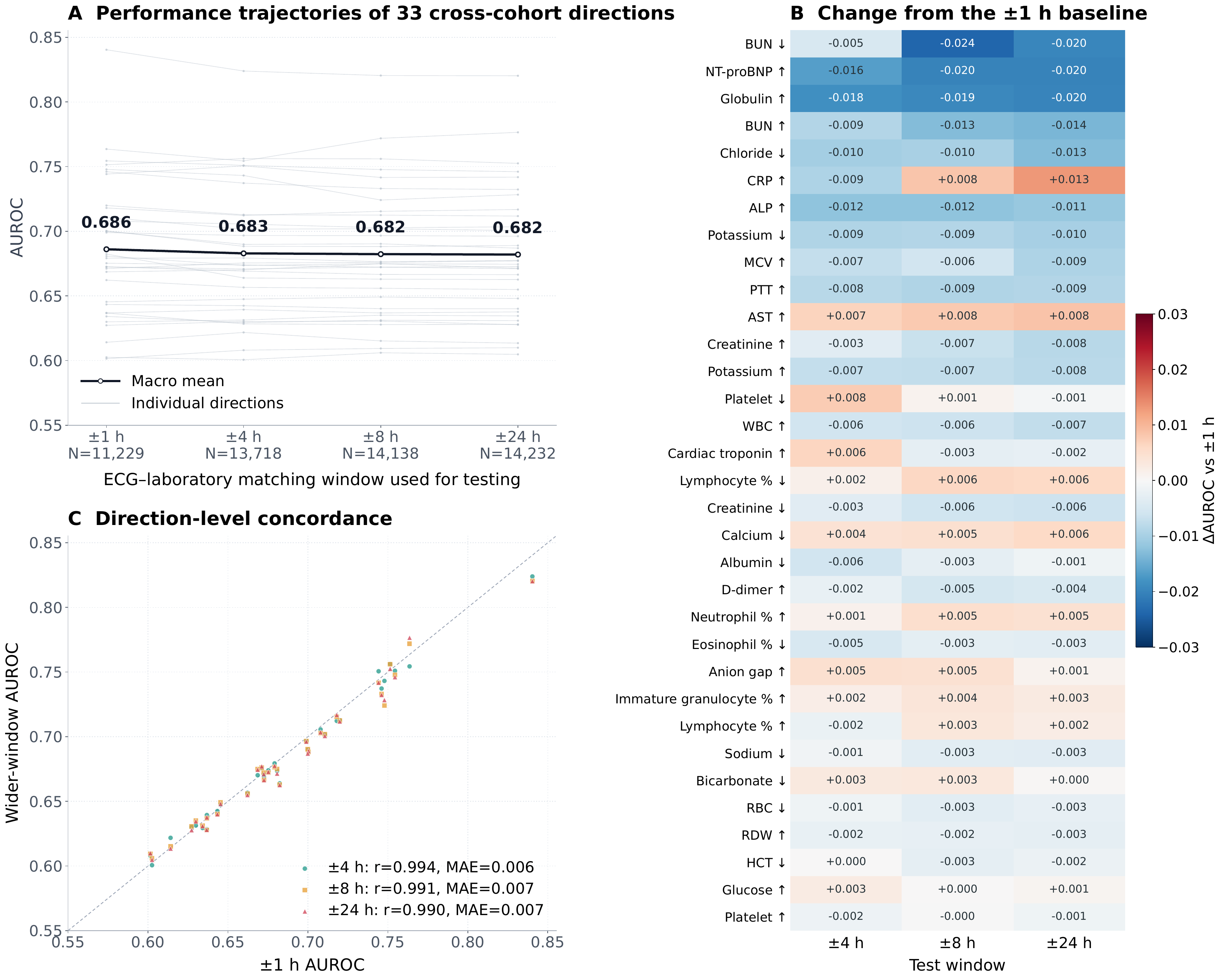}
\caption{\textbf{Sensitivity to the ECG--laboratory matching
window.}
The seed-42 checkpoint trained with a $\pm$1-hour window was
evaluated after the bidirectional MC-MED test matching window was
widened to $\pm$4, $\pm$8, and $\pm$24 hours. (A) Sample-level AUROC
trajectories for the 33 shortlisted directions and their macro mean.
(B) Direction-level change from the one-hour reference.
(C) Direction-level AUROCs compared across windows. No retraining or
threshold selection was performed. The analysis tests sensitivity to
the matching rule; it does not test prediction of future
abnormalities.}
\label{fig:temporal}
\end{figure}

We next assessed sensitivity to ECG--laboratory matching while
holding the seed-42 checkpoint fixed. This model had been trained
with a $\pm$1-hour window; only the matching window in the MC-MED
test partition was widened. Across the 33 shortlisted directions,
the sample-level macro AUROC was 0.686
at $\pm$1 hour, 0.683 at $\pm$4 hours,
0.682 at $\pm$8 hours, and
0.682 at $\pm$24 hours
(Figs.~\ref{fig:architecture}d and \ref{fig:temporal}). Over the
same windows, the number of matched test samples increased from
11{,}229 to 14{,}232. Direction-level AUROCs remained closely
correlated with the one-hour estimates, with Pearson correlations
of 0.994,
0.991, and 0.990
at $\pm$4, $\pm$8, and $\pm$24 hours, respectively. The mean absolute change in AUROC was 0.006--0.007,
and the largest change for an individual direction was 0.024.

\subsection*{Model predictions reproduce and amplify the coupling
among shortlisted abnormalities}
\label{sec:cooccurrence}

We next asked whether relationships among the model's predictions
matched those among the observed laboratory labels. As a contextual
reference, we summarized co-occurrence in the available laboratory
histories of 31{,}826 MC-MED and 115{,}921 MIMIC-IV patients
(Supplementary Fig.~\ref{fig:cooccurrence}). These associations
largely reflected established relationships among laboratory tests,
such as sodium with chloride and hematocrit with red cell count;
they do not by themselves provide evidence of ECG-specific
information. We therefore focused on the comparison between
observed-label and prediction coupling rather than on these known
laboratory associations.

\begin{figure}[p]
\centering
\includegraphics[width=\linewidth,height=0.76\textheight,keepaspectratio]
{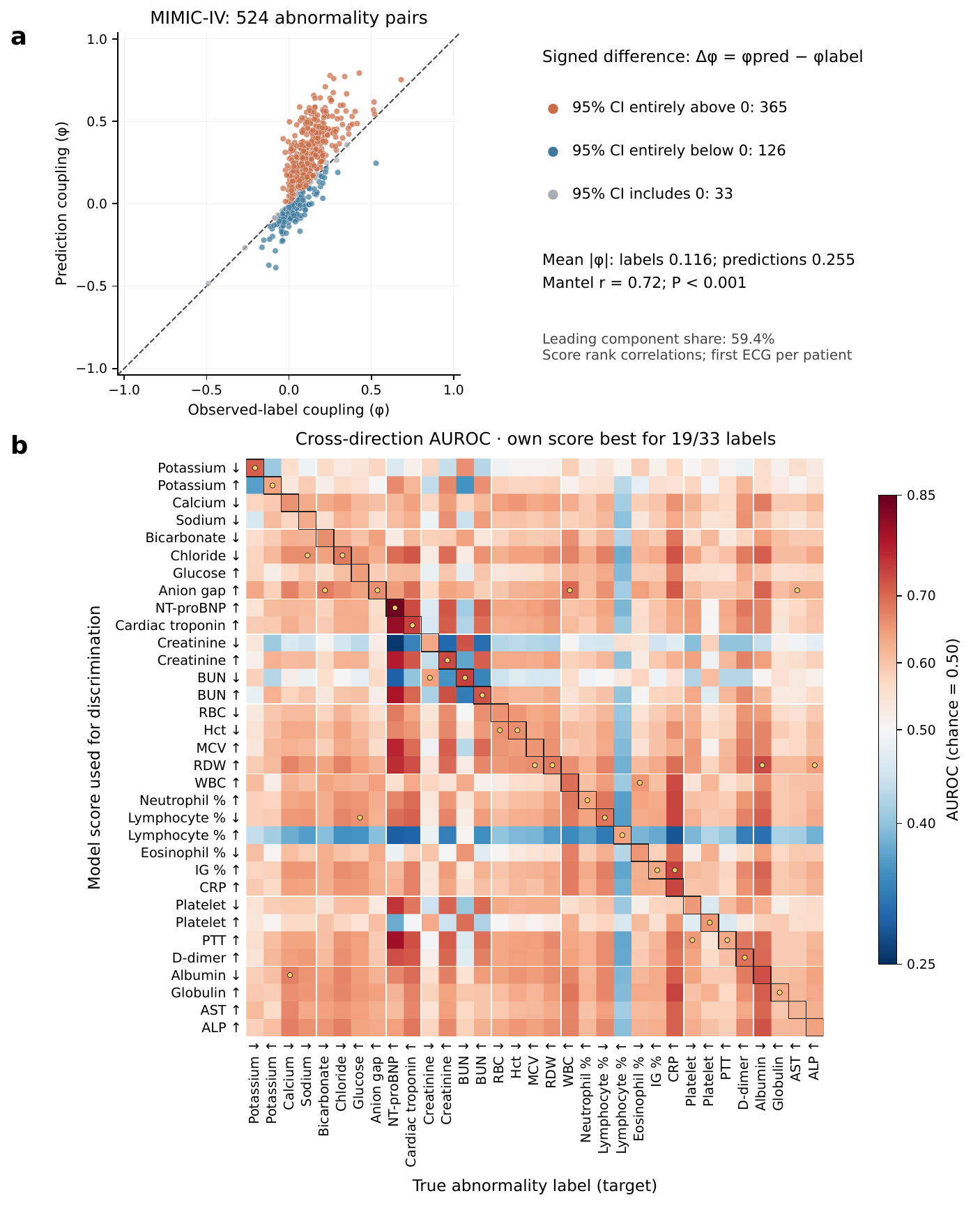}
\captionsetup{font=small}
\caption{\textbf{Coupling and direction specificity of ECG-derived
laboratory predictions in MIMIC-IV.}
\textbf{(a)} Observed-label versus prediction $\varphi$ for 524
evaluable pairs among the 33 shortlisted directions, using the same
earliest co-observed ECG per patient. Predictions are binarized at
observed prevalence. The dashed line is $y=x$. Colors indicate
whether the 200-resample bootstrap 95\% CI for
$\Delta\varphi=\varphi_{\mathrm{pred}}-\varphi_{\mathrm{label}}$
is above zero, below zero, or includes zero. A positive
$\Delta\varphi$ does not necessarily mean stronger $|\varphi|$.
The leading-component share uses score rank correlations on each
patient's first ECG.
\textbf{(b)} Rows are model scores; columns are true abnormality
labels. Each column uses its target's earliest observed ECG per
patient. Outlined cells show own-score AUROCs; gold dots mark column
maxima, including 19 of 33 diagonal cells. Colors are centered at
AUROC 0.50; below-chance scores are not reversed. Best-score
comparisons are descriptive, not significance tests. Both panels
use the seed-42 model.}
\label{fig:coupling_specificity}
\end{figure}

To compare these relationships with the model, we used the same ECG
for labels and predictions and one ECG per patient (Methods). For
each pair, we computed the $\varphi$ coefficient between the observed
labels and between the predictions after binarizing each score at
the observed prevalence. In MIMIC-IV, the two sets of pairwise
coefficients were correlated (Mantel $r=0.72$, $P<0.001$; MC-MED
$r=0.70$, $P<0.001$), so the model reproduced the overall structure
of the observed abnormalities. Its predictions were, however, more
tightly coupled than the labels. The mean absolute $\varphi$ was 0.25
for predictions and 0.12 for labels in MIMIC-IV (0.26 and 0.12 in
MC-MED). The signed difference,
$\Delta\varphi=\varphi_{\mathrm{pred}}-\varphi_{\mathrm{label}}$,
had a bootstrap 95\% CI entirely above zero in 365 of 524 evaluable
pairs and entirely below zero in 126
(Fig.~\ref{fig:coupling_specificity}a). These counts describe changes
in signed association, not changes in absolute coupling strength.
The largest positive differences involved pairs that were
nearly independent in the labels, such as elevated red cell
distribution width with elevated anion gap ($\varphi=0.07$ for labels
and 0.59 for predictions). Conversely, low sodium and low chloride
were less coupled in the predictions (0.25) than in the labels
(0.53). In the pairwise $\varphi$ matrices, the leading component accounted
for 30\% of the eigenvalue sum for predictions and 16\% for labels,
and the participation ratio, an index of how many components share
the variance, was 8.3 and 19.2, respectively. Across the first ECGs
of all 115{,}921 MIMIC-IV patients, a single component accounted for
59\% of the variance in the 33 prediction scores (participation
ratio 2.5).

Many directions nevertheless carried specific information. In
MIMIC-IV, a direction's own score gave the highest AUROC for its
label in 19 of the 33 directions (24 in MC-MED;
Fig.~\ref{fig:coupling_specificity}b). The margin over the
best alternative score was largest for elevated lymphocyte
percentage (0.081), low potassium (0.074), elevated NT-proBNP
(0.033, relative to the troponin score), and low BUN (0.020). For low
sodium, low bicarbonate, and low calcium, the score of a related
direction (low chloride, elevated anion gap, and low albumin,
respectively) discriminated the label better than the direction's own
score. Within strata of a strongly associated test, discrimination
was largely retained for elevated cardiac troponin (AUROC 0.713 and
0.672 among patients without and with elevated creatinine) and
elevated creatinine (0.661 and 0.632 within strata of elevated BUN),
but low sodium fell to 0.557 among patients with low chloride and
low calcium to 0.550 among patients with low albumin. The model
therefore reproduced the structure of the observed abnormalities
but compressed much of it onto one shared component, consistent with
a common signal of illness severity, while retaining
analyte-specific signal for a subset of directions.

\subsection*{ECG-mGPS and hybrid ECG-SIRS are associated with mortality}
\label{sec:clinical-scores}

\begin{figure}[htbp]
\centering
\includegraphics[width=\linewidth]
{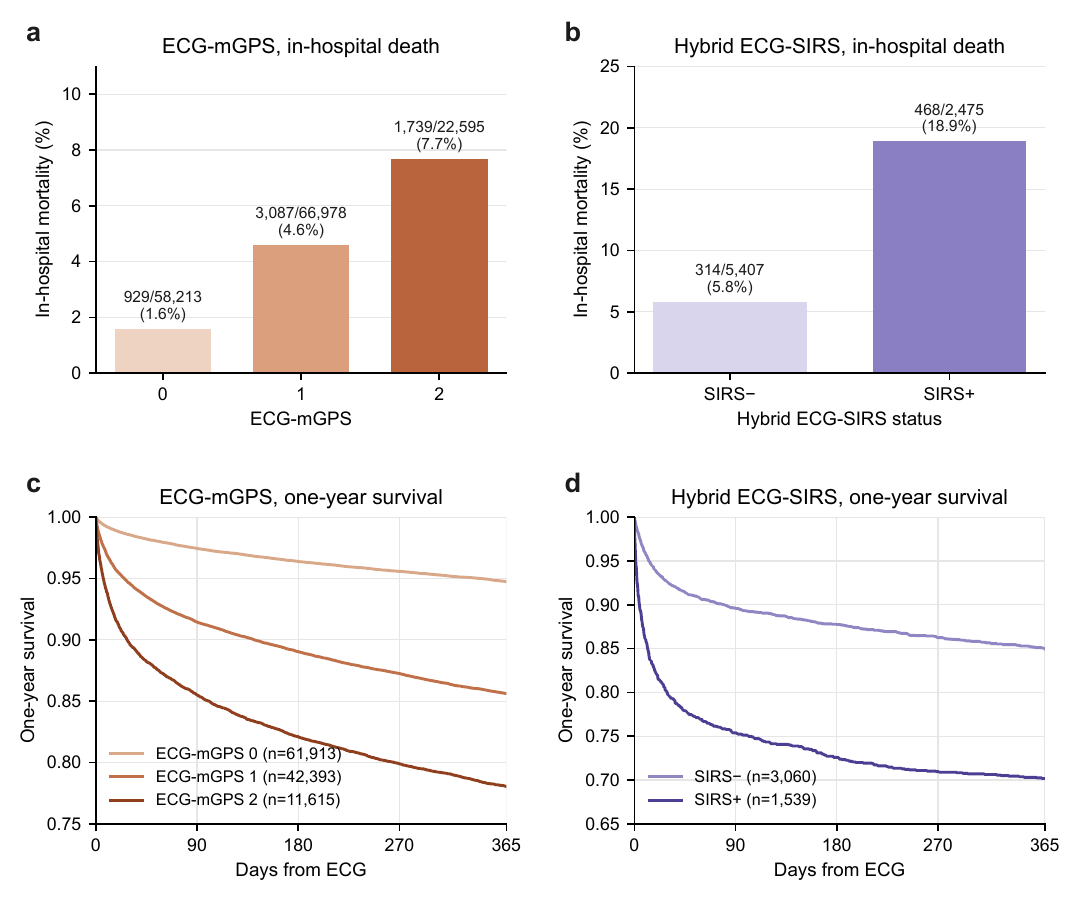}
\caption{\textbf{ECG-mGPS, hybrid ECG-SIRS, and mortality
in MIMIC-IV.}
\textbf{(a)} In-hospital mortality by hard ECG-mGPS score (0--2),
using the first ECG from each matched admission.
\textbf{(b)} In-hospital mortality by hybrid ECG-SIRS status in the
ICU-vitals complete-case cohort, using the first ECG from each
matched admission. SIRS-negative admissions (hard score $<2$) are
compared with SIRS-positive admissions (hard score $\ge2$).
\textbf{(c)} Kaplan--Meier one-year survival from the first ECG per
patient, stratified by hard ECG-mGPS score.
\textbf{(d)} Kaplan--Meier one-year survival from the first ECG per
patient in the complete-case cohort, stratified by hybrid ECG-SIRS
status. ECG-mGPS and hybrid ECG-SIRS were evaluated in different
eligible populations.}
\label{fig:clinical_scores_mortality}
\end{figure}

We extended the laboratory screen to an exploratory assessment of
prognostic relevance by incorporating model outputs into ECG-mGPS
and hybrid ECG-SIRS, whose laboratory components (CRP and albumin
for mGPS, WBC for SIRS) belong to the prespecified panel. We examined their associations with in-hospital and one-year mortality in MIMIC-IV
(Figs.~\ref{fig:architecture}d and
\ref{fig:clinical_scores_mortality}).

ECG-mGPS combined binary classifications of elevated CRP and low
albumin, using thresholds fixed in the MC-MED validation partition,
to assign scores of 0, 1, or 2. Hybrid ECG-SIRS combined the
model-based classification of abnormal WBC with ECG-derived heart
rate and intensive care unit (ICU) temperature and respiratory
rate recorded within $\pm$1 hour of the ECG. ECG-mGPS was evaluated
in the broader admission-matched cohort, whereas hybrid ECG-SIRS
required all four components and was evaluated in a complete-case
cohort with available ICU vital signs.

For ECG-mGPS, the in-hospital analysis included 147{,}786 matched
admissions, retaining the first ECG from each admission.
In-hospital mortality increased from 1.6\%
(929/58{,}213) at score 0 to
4.6\% (3{,}087/66{,}978) at score 1 and
7.7\% (1{,}739/22{,}595) at score 2
(Fig.~\ref{fig:clinical_scores_mortality}a).
Compared with score 0, the age- and sex-adjusted odds ratios were
2.65 (95\% CI 2.46--2.86) for score 1 and 5.15
(95\% CI 4.75--5.59) for score 2. The patient-level one-year
analysis included 115{,}921 patients, with an overall mortality
of 10.3\%. One-year mortality was
5.2\%, 14.4\%, and
22.0\% for scores 0, 1, and 2, respectively
(log-rank $P<0.001$;
Fig.~\ref{fig:clinical_scores_mortality}c).
The corresponding age- and sex-adjusted hazard ratios relative
to score 0 were 2.23 (95\% CI 2.14--2.33) and 4.01
(95\% CI 3.81--4.23).

For hybrid ECG-SIRS, the in-hospital analysis included 7{,}882
admissions whose first matched ECG had all four components
available. Patients with multiple admissions could
contribute more than one admission. Overall in-hospital mortality
was 9.9\% (782/7{,}882), with
mortality of 5.8\% (314/5{,}407) in the
SIRS-negative group and 18.9\%
(468/2{,}475) in the SIRS-positive group
(Fig.~\ref{fig:clinical_scores_mortality}b).
The age- and sex-adjusted odds ratio for SIRS-positive versus
SIRS-negative status was 4.14 (95\% CI 3.55--4.83).
The patient-level one-year analysis included 4{,}599 patients,
with an overall mortality of 20.0\%.
One-year mortality was 15.1\%
(461/3{,}060) in the SIRS-negative group
and 29.8\% (459/1{,}539) in the
SIRS-positive group
(log-rank $P<0.001$;
Fig.~\ref{fig:clinical_scores_mortality}d).
The age- and sex-adjusted hazard ratio was 2.52
(95\% CI 2.21--2.87).

We also compared the ECG-based scores with their laboratory-defined
counterparts in all matched ECGs with the required tests. ECG-mGPS
matched laboratory mGPS in 611 of 1{,}214 ECGs (50\%;
quadratic-weighted $\kappa=0.36$). Laboratory SIRS shared the
temperature, respiratory rate, and heart rate of hybrid ECG-SIRS and
differed only in using the measured WBC count. In 10{,}512
complete-case ECGs with WBC measured, SIRS-positive status agreed in
82\% and the quadratic-weighted $\kappa$ for the 0--4 score was 0.79.
These figures mostly reflect the shared components. The model-based
and measured WBC components agreed in 53\% of these ECGs
($\kappa=0.07$), and the model classified WBC as abnormal in 28\%,
compared with 52\% by measurement.

\FloatBarrier


\section*{DISCUSSION}
\label{sec:discussion}

A single-lead ECG model fine-tuned in one health system met
the same screening criteria for 33 laboratory abnormality
directions in a second, with a mean patient-first AUROC of 0.692
in MC-MED and 0.675 in MIMIC-IV. Discrimination was stable across
training seeds and matching windows, but much of it was shared
across directions, and only a subset of directions kept
discrimination beyond that of correlated tests.

Previous AI-ECG studies have demonstrated the feasibility of
detecting individual laboratory abnormalities, including abnormal
potassium from limited-lead and smartwatch
ECGs\cite{galloway2019development,chiu2024serum}.
CardioLab extended this work to a broader laboratory panel using
12-lead recordings\cite{alcaraz2025abnormality}.
Our study examined a prespecified panel of 50 tests using
single-lead waveforms alone and applied the same patient-first
screening criteria in two health systems with different ECG
acquisition settings. Compared with our earlier AnyECG-Lab
preprint\cite{xiao2025anyecg}, the present study adds
frozen-model external evaluation and an explicit cross-cohort
intersection screen. The resulting shortlist indicates which laboratory abnormalities
to prioritize in single-lead ECG studies.

Aggregate task-level discrimination changed little across three
training seeds, and discrimination for the fixed shortlist changed
little when the ECG--laboratory matching window was widened from
$\pm$1 to $\pm$24 hours. The matching-window analysis concerns how
ECGs were paired with laboratory measurements; evaluating the
prediction of future abnormalities would require ECG-first indexing
and a predefined prediction horizon.

The shortlisted abnormalities span cardiac, renal, electrolyte,
hematologic, and inflammatory processes. Such disturbances may
affect ECG waveforms through electrophysiological changes,
oxygen delivery, autonomic tone, and overall illness severity.
The observed-label network mainly reflected established relationships
among laboratory tests and served as a contextual reference
(Supplementary Fig.~\ref{fig:cooccurrence}), not evidence of novel
or ECG-specific associations. More informative for interpreting the
ECG model was the comparison between observed-label and prediction
coupling: predictions were about twice as tightly coupled as the
labels, and across all patients a single component accounted for
most of their variance. Part of the
discrimination for many directions may therefore reflect a common
signal of illness severity rather than analyte-specific physiology.
Low potassium, elevated lymphocyte percentage, elevated NT-proBNP
and cardiac troponin, elevated creatinine, and elevated and low BUN
kept discrimination beyond that of correlated tests, whereas low sodium, low bicarbonate, and low
calcium did not. Directions in the first group are the more
promising targets for analyte-specific study.

No thyroid or lipid direction met the shortlist criteria. In MC-MED,
these tests were rarely ordered in the emergency setting, so these
directions could not be adequately trained or evaluated, and their
absence from the shortlist should not be read as absence of ECG
information. In MIMIC-IV, where sample sizes were adequate, lipid
directions showed AUROCs of 0.56--0.60, consistent with the lack of a
direct electrophysiological pathway, whereas elevated free thyroxine
reached 0.673, consistent with the chronotropic effects of
hyperthyroidism and warranting evaluation in a cohort with routine
thyroid testing.

ECG-mGPS categories were associated with in-hospital and
one-year mortality in the broader eligible population, and
hybrid ECG-SIRS showed mortality differences in the smaller
cohort with available ICU vital signs. These associations should be
read together with the coupling analysis. ECG-mGPS agreed only
moderately with laboratory mGPS ($\kappa=0.36$), and for both of its
components the scores of other directions discriminated the label
about as well as the component's own score. Its association with
mortality may therefore reflect overall illness severity captured by
the ECG as much as inflammation or nutritional status. The higher
$\kappa$ for hybrid ECG-SIRS came from the three components it shared
with laboratory SIRS. In these ICU patients the model-based WBC
component agreed with the measured count little better than chance
($\kappa=0.07$), so the mortality association of hybrid ECG-SIRS
probably rests mainly on the vital signs and ECG-derived heart rate.
Whether the model-based WBC component adds information beyond them
remains to be tested.

Several aspects of the design affect interpretation. Both cohorts
contributed to selection of the 33-direction shortlist, so its
performance estimates are descriptive and require confirmation
in an additional cohort. The external troponin label was defined
after troponin I proved too sparse in MIMIC-IV, and the grouping of
tasks into heads reflects a mechanistic hypothesis that was not
compared with alternative groupings. The seed analysis assessed aggregate
discrimination rather than the stability of individual shortlist
membership. Performance also varied across abnormality directions
and between cohorts. Differences in populations, laboratory
assays, reference intervals, ordering practices, and ECG
acquisition may contribute to these shifts; probability calibration
and decision thresholds therefore require evaluation in the
intended use setting.

Laboratory measurements were obtained for clinical reasons and
were missing not at random. Observation masks prevented unmeasured
tests from being treated as normal, but did not remove selection
associated with testing. The co-occurrence analysis was unadjusted
and summarized abnormalities across each patient's available
history, making its associations dependent on testing frequency
and observation duration. Although the prediction model received
no clinical covariates, ECG waveforms contain information related
to age, heart rate, and illness severity, which may contribute
to laboratory discrimination. The mortality analyses adjusted for age and sex but may remain confounded by disease burden and care setting, and the hybrid ECG-SIRS cohort was selected by ICU-vital-sign availability. Because the CRP and albumin thresholds were matched to prevalences among patients in whom these tests were ordered, about 60\% of MIMIC-IV admissions were classified by the model as having elevated CRP, so ECG-mGPS categories are not directly
comparable with laboratory-defined mGPS. The WBC threshold was miscalibrated in the opposite direction: among ICU ECGs with WBC measured, the model classified 28\% as abnormal, whereas 52\% were abnormal by measurement. External evaluation was limited to one US database, with no evaluation on wearable recordings or assessment of subgroup performance and fairness.

Further work should focus on candidate abnormalities and scores
with a defined clinical use. This includes reviewing laboratory
mappings and thresholds, evaluating calibration and clinically
relevant operating points, and testing whether ECG-based estimates
add information beyond routine clinical variables.
Prospective workflow studies following the Developmental and
Exploratory Clinical Investigations of Decision support systems
driven by Artificial Intelligence
(DECIDE-AI) framework\cite{vasey2022reporting}
would then establish whether this information improves clinical
decisions.


\section*{METHODS}
\label{sec:methods}

\subsection*{Cohort construction and ECG--laboratory matching}
\label{sec:datasets}

MC-MED is a de-identified waveform and laboratory database of adult
emergency-department visits to Stanford Health Care, Stanford,
California, USA, collected between September 2020 and September
2022\cite{PhysioNet-mc-med-1.0.1,kansal2025mc}. Continuous bedside
monitoring provided single-lead (lead II) ECG waveforms. The source
laboratory table contained 5{,}706{,}470 records. We kept a record if
its laboratory test belonged to the predefined panel, its numerical
result could be parsed, its unit matched the task definition, and its
encounter could be linked to a patient identifier. For the final
50-test panel, this left 2{,}740{,}304 laboratory records from
58{,}454 label-eligible patients.

Patients were assigned to patient-disjoint training, validation, and test partitions in an approximate 7:2:1 ratio. To ensure a deterministic and reproducible assignment, patients were ordered by their earliest eligible laboratory record as recorded in the de-identified MC-MED dataset. The assignment was made on the broader candidate panel and kept
unchanged when the panel was finalized. Before ECG matching, the
partitions contained 40{,}919, 11{,}690, and 5{,}845 patients with at
least one record in the final panel; no patient appeared in more than
one partition. After waveform availability and
quality filters, the analysis included 22{,}275 training, 6{,}364
validation, and 3{,}187 test patients, contributing 80{,}590,
23{,}586, and 11{,}229 model samples.

MIMIC-IV is a de-identified hospital database covering critical and
emergency care\cite{johnson2023mimic}. Laboratory events from the
hospital module, version 3.1, were linked to the MIMIC-IV-ECG
Diagnostic Electrocardiogram Matched Subset, version 1.0, which
contains 10-second, 12-lead diagnostic ECGs sampled at
500\,Hz\cite{gow2023mimic}. MC-MED laboratory tests were mapped to
MIMIC-IV \texttt{d\_labitems} entries after the analyte meaning and
units were checked. The external analysis included 47 mapped tasks,
300{,}035 ECGs, and 115{,}921 patients. Three laboratory tests could not be mapped: transferrin saturation, non-high-density lipoprotein (non-HDL) cholesterol, and procalcitonin. Label thresholds defined in MC-MED were used without
external tuning.

Troponin I was rarely measured in MIMIC-IV, where troponin T was the
routine cardiac troponin assay. The external label for the troponin
task therefore combined both assays, each judged against its own
threshold: troponin I above 0.04\,ng/mL, the panel threshold, and
troponin T above 0.01\,ng/mL, the laboratory upper reference limit.
When both were available for the same ECG, the result closest in time
was used. This definition was adopted after troponin I proved too
sparse in MIMIC-IV for evaluation, so external estimates for the
troponin direction should be read with that in mind. In MIMIC-IV,
results reported as text rather than numbers were handled with a
single rule for all tests. A numeric value was used when available.
Otherwise, a result reported with a leading comparator, such as
``$<$0.01'', was read as an interval, and
a label was assigned only when the whole interval fell within one
class, with values at a reference limit counted as normal. Intervals
spanning more than one class and other text results were treated as
missing; no point value was imputed. This rule recovered labels in 24
of the 47 mapped tests, for 295{,}081 laboratory records before ECG
matching, of which 268{,}009 were troponin T results.

The clinical-score analyses also used hospital admissions, patient
vital status, and ICU \texttt{chartevents} from MIMIC-IV version 3.1.
Demographics and coded diagnoses in Table~\ref{tab:population} were
taken from the version 2.2 hospital module.
Patient-level race and ethnicity in Table~\ref{tab:population} were
taken as the most frequent known category across hospital admissions
and MIMIC-IV-ED version 2.2 stay records\cite{johnson2021mimic}.
Emergency-department stay records were used only for this
demographic description. Comorbidities in
Table~\ref{tab:population} use the same patients. In MC-MED, a
condition was counted when an ICD-9 or ICD-10 past-history code was
noted on or before the patient's earliest model ECG. In MIMIC-IV, a
condition was counted when the same code groups appeared on a
hospital admission or emergency-department stay that began at or
before that ECG. The groups cover hypertension, diabetes,
dyslipidemia, coronary artery disease, heart failure, atrial
fibrillation or flutter, chronic kidney disease, chronic liver
disease, chronic pulmonary disease, obesity, anemia, and thyroid
disease. A patient with no matching code was counted as not having
the condition.

In MC-MED, laboratory order times were matched to continuous
single-lead waveform intervals within a bidirectional $\pm$1-hour
window. When several waveform segments were eligible, we kept the one
with the longest overlap, and segments shorter than 10 seconds were
excluded. In MIMIC-IV, each laboratory event was matched to the
nearest ECG from the same patient within $\pm$1 hour, and a single
lead (lead II) was extracted from the 12-lead recording.

Each model input was a 10-second single-lead signal sampled at
500\,Hz (5{,}000 samples). Signals were linearly resampled when
necessary and padded or truncated to this length. Preprocessing
consisted of a 50-Hz notch filter (quality factor 30), a fourth-order
0.67--40-Hz Butterworth band-pass filter, median-filter baseline
removal, and $z$-score normalization of each waveform. Waveforms that
could not be decoded or failed the minimum-duration check were
excluded.

The primary $\pm$1-hour rule was intended to capture
near-contemporaneous abnormalities. In sensitivity analyses, the
MC-MED test matching window was widened to $\pm$4, $\pm$8, and
$\pm$24 hours, and the model trained with the $\pm$1-hour window was
kept unchanged.

\subsection*{Study design and laboratory endpoint definitions}
\label{sec:design}

This retrospective study combined model development in MC-MED with
frozen-model evaluation in MIMIC-IV. MC-MED was used to define tasks,
fine-tune the model, select the checkpoint, set clinical-score
classification thresholds, and perform internal testing. MIMIC-IV was
used for external evaluation, cross-cohort shortlist selection, the
laboratory association and coupling analyses, and the exploratory
clinical-score analyses. No MIMIC-IV samples were used for parameter updates,
checkpoint selection, hyperparameter selection, or clinical-score
threshold selection.

The screen aimed to identify laboratory abnormality directions that
met common statistical, sample-size, and discrimination criteria in
both the MC-MED test partition and MIMIC-IV. Because both cohorts
were used to select the final set, the 33-direction shortlist is
descriptive and was not a prespecified confirmatory panel.

Each task used unit-specific reference thresholds defined before
model training. The study-defined LOW and HIGH thresholds, units,
and cohort-specific exceptions are listed in Supplementary
Table~\ref{tab:thresholds}. Tasks with a non-zero lower threshold had three
classes: LOW (class 0), NORMAL (class 1), and HIGH (class 2). Tasks
without a lower reference limit had two: NORMAL (class 0) and HIGH
(class 1). Before training, and without reference to any predictive
performance, we finalized the panel from an earlier, broader
candidate set. The removed assays included ionized and corrected
calcium, calculated osmolality, osmolal gap, total protein, total
cholesterol, unsaturated iron-binding capacity, and serum ethanol,
and D-dimer measurements were merged into a single task. The
resulting prespecified panel contained 50 laboratory tests.

Cut-points for three laboratory tests were aligned with the clinical-score definitions used later, with 1.0\,mg/dL for CRP HIGH, 3.5\,g/dL for albumin LOW, and 4 and 12\,K/$\mu$L for WBC LOW and HIGH. The final panel comprised 39 ternary and 11 binary tasks. Each
ternary task contributed two one-versus-rest abnormality directions
and each binary task one, giving 89 directions.

Each sample carried a task-specific observation mask. Unmeasured
laboratory tests received a mask value of zero and were excluded from
the corresponding loss and metrics instead of being coded as normal.
When a laboratory order contained several results for the same task,
the most recently reported result was kept.

\subsection*{Fine-tuning ECGFounder for laboratory-wide screening}
\label{sec:model}

The backbone was initialized from the published single-lead
ECGFounder checkpoint. Its shared 1,024-dimensional representation
fed ten classification heads, and all trainable parameters were
fine-tuned. The heads grouped tasks by the presumed pathway through
which an abnormality could affect the ECG: electrolytes with direct
effects on the myocardial action potential (potassium, calcium,
magnesium, and sodium); acid--base and metabolic status (bicarbonate,
chloride, lactate, glucose, serum osmolality, and anion gap);
myocardial load and injury (NT-proBNP, troponin, creatine kinase, and
lactate dehydrogenase); renal function and minerals (creatinine, BUN,
and phosphorus); erythroid and iron indices (red blood cell count,
hematocrit, mean corpuscular volume, red cell distribution width,
iron, total iron-binding capacity, ferritin, and transferrin
saturation); inflammation and immunity (white blood cell count and
differential percentages, immature granulocytes, CRP, erythrocyte
sedimentation rate, and procalcitonin); hemostasis (platelet count,
partial thromboplastin time, and D-dimer); hepatic synthesis and
gastrointestinal function (albumin, globulin, aspartate
aminotransferase [AST], alkaline phosphatase, ammonia, and lipase); thyroid function
(thyroid-stimulating hormone and free thyroxine); and lipids (HDL,
LDL, and non-HDL cholesterol, and triglycerides). The grouping states
a hypothesis about mechanism; it does not constrain what each head
learns.

Training used masked cross-entropy loss, so only observed labels
contributed to optimization. Class weights were inversely
proportional to class frequency and capped at 10. Losses were
averaged within each head, and each head's loss was weighted by its
number of tasks, so that every task contributed approximately equally
to the final objective.

Models were trained with the Adam optimizer (initial learning rate
$10^{-4}$, weight decay $10^{-5}$) and a batch size of 64. The
primary training seed was 42, and training was scheduled for at most
15 epochs. The learning rate was multiplied by 0.1 after two epochs
without improvement in the validation mean task-level abnormal-class
AUROC, and training stopped when the learning rate fell below
$10^{-5}$.

Checkpoints were selected on the MC-MED validation partition only,
using the mean task-level abnormal-class AUROC. For ternary tasks,
the LOW and HIGH one-versus-rest AUROCs were macro-averaged; for
binary tasks, the HIGH-class AUROC was used. The held-out test
partition was evaluated only after the checkpoint had been chosen.

\subsection*{Patient-first evaluation and cross-cohort screening}
\label{sec:evaluation}

The primary performance measure was AUROC. To limit dependence among
repeated observations from the same patient, the primary
direction-level analysis kept only the earliest observed record for
each patient and task (patient-first analysis). For each direction,
the target abnormal class was treated as positive and all other
observed classes of that task as negative. The percentage beside each patient count in Tables~\ref{tab:cross_cohort} and \ref{tab:s1} is the positive rate, defined as positive patients divided by the evaluable patients in that direction.

Prediction scores of positive and negative patients were compared
with a two-sided Mann--Whitney $U$ test. Both cohorts used a
family-wise significance level of 0.01 over the same 89-direction
family, so the Bonferroni criterion required a raw $P$ value below
\[
\frac{0.01}{89}\approx 1.12\times10^{-4}.
\]
A direction entered the shortlist only if it met all of the following criteria independently in both cohorts, namely the Bonferroni criterion, at least 30 positive patients, at least 30 negative patients, more than 100 observed patients in total, and AUROC $>0.60$.

Patient-first AUROC confidence intervals were estimated from 1,000
stratified bootstrap iterations with seed 42, sampling positive and
negative patients separately with replacement. The 2.5th and 97.5th
percentiles defined the 95\% confidence interval.
These intervals are direction-specific and are not simultaneous
confidence intervals adjusted for testing the full panel.

For external evaluation, the complete MC-MED checkpoint was loaded
without further training, and unmapped tasks remained masked.
Evaluable directions were assessed with the same patient-first rules,
and the multiplicity denominator remained 89. Directions that met all
criteria in both cohorts formed the final shortlist.

Fine-tuning was repeated with seeds 123 and 456 on the same patient
partition, and each validation-selected checkpoint was evaluated on
the same 47 mapped MIMIC-IV tasks. These results were summarized by
the sample-level task-macro abnormal-class AUROC, using all observed
matched samples without patient-first reduction. Within each task,
evaluable abnormal-class AUROCs were averaged first; these task-level
values were then averaged with equal weight across the 50 MC-MED
tasks or the 47 mapped MIMIC-IV tasks. This is a different metric
from the patient-first direction-level AUROC used for shortlist
selection.

For matching-window sensitivity, the seed-42 checkpoint trained with
the $\pm$1-hour window was held fixed, and only the MC-MED test
matching table was replaced by the $\pm$4-, $\pm$8-, or $\pm$24-hour
version. The same 33 directions were analyzed without further
selection. We compared sample-level macro AUROC, the mean and maximum
absolute change in direction-level AUROC, and Pearson correlations
with the one-hour reference.

\subsection*{Co-occurrence analysis of shortlisted abnormalities}
\label{sec:network}

The network used observed laboratory labels for the 33 shortlisted
directions. For each direction, a patient was assigned status 1 if
the abnormality had occurred at least once in the available history,
status 0 if the test had been measured but that direction never
occurred, and missing if the test had never been measured.

For each pair of directions, the analysis included patients in whom
both tests had been measured; pairs with fewer than 50 such patients
were not tested. For each pair, we built a $2\times2$ table,
calculated the signed phi coefficient and odds ratio, and tested
independence with a two-sided Fisher exact test.

$P$ values were adjusted separately in each cohort with the
Benjamini--Hochberg procedure. A shared edge required $q\le0.05$ in
both cohorts and phi coefficients of the same sign. The figure shows
the 50 shared edges with the largest mean absolute phi coefficients.

To compare label and prediction coupling, each pair of directions was
analyzed in patients whose matched ECG had both tests observed,
keeping the earliest such ECG per patient, so that labels and
predictions came from the same ECG. The label coupling was the $\varphi$
coefficient between the two binary labels. For the prediction
coupling, each direction's predicted probability was binarized at
the quantile that reproduced the observed prevalence in that sample,
so that both coefficients had the same marginal rates, and $\varphi$
was computed between the binarized predictions. Signed differences
$\Delta\varphi=\varphi_{\mathrm{pred}}-\varphi_{\mathrm{label}}$
were summarized with 200 patient bootstrap resamples; a positive
difference is not necessarily an increase in $|\varphi|$.
Agreement between
the label and prediction coupling matrices was assessed with a Mantel
test with 9{,}999 permutations. Dimensionality was summarized by the
leading-component share of the nonnegative eigenvalue sum and by the
participation ratio, $(\sum_k\lambda_k)^2/\sum_k\lambda_k^2$, of the
eigenvalues $\lambda_k$ of each coupling matrix; for predictions, this
was also computed from the Spearman correlations of the scores on
the first ECG of every patient. Each pair used its own patient
subset, pairs that could not be evaluated were set to zero, and
negative eigenvalues were truncated at zero, so these summaries
describe correlation structure rather than a count of independent
biological processes. Direction specificity was assessed by
the AUROC of every direction's score for every direction's label,
using the earliest observed ECG per patient for that label, and by
the AUROC of each score within strata of a strongly associated label
($|\varphi|\ge0.3$), using each patient's earliest ECG on which both
tests had been observed. These analyses used the seed-42 model.

\subsection*{Construction of ECG-based scores and mortality analyses}
\label{sec:score-methods}

To examine the prognostic relevance of the laboratory predictions,
we constructed ECG-mGPS and hybrid ECG-SIRS from model outputs
and the clinical measurements required by each score.
Classification thresholds for the predicted laboratory components
were selected in the MC-MED validation partition by matching
predicted positive rates to observed prevalences and were kept
fixed for the MIMIC-IV analyses.

We used two established, rule-based scores as prespecified
templates, because their thresholds are fixed a priori and require
no model fitting. The first was the modified Glasgow Prognostic
Score (mGPS). It was developed and validated mainly in oncology.
However, its components independently predict all-cause,
cardiovascular, and cerebrovascular mortality in a large unselected
cohort of hospital and community patients\cite{proctor2015systemic},
and the score has been associated with mortality in non-malignant
conditions, including heart failure\cite{cho2020inflammation}. The second was
the systemic inflammatory response syndrome (SIRS)
criteria\cite{bone1992definitions}. These combine one laboratory
component, the white blood cell count, with three vital-sign
components. This structure allows a single model-derived component
to be substituted while the others are measured directly. SIRS is
no longer used to define sepsis\cite{singer2016third}, so we used it
only as a composite of physiological derangement. Meeting at least
two criteria has been associated with higher in-hospital mortality
in general ward patients\cite{churpek2015incidence}, and each
additional criterion has been associated with higher adjusted
mortality in ICU patients\cite{kaukonen2015systemic}.

ECG-mGPS followed the asymmetric structure of the modified Glasgow
Prognostic Score\cite{forrest2003evaluation,mcmillan2007evaluation}.
The score was 0 when CRP was not classified as elevated, 1 when
CRP was classified as elevated but albumin was not classified as
low, and 2 when both elevated CRP and low albumin were predicted.
Mortality analyses used these categorical scores.

Hybrid ECG-SIRS combined four binary
components\cite{bone1992definitions}.
Temperature $>38$ or $<36\,^{\circ}\mathrm{C}$ and respiratory
rate $>20\,\mathrm{min}^{-1}$ were obtained from the nearest
ICU \texttt{chartevents} entries within $\pm$1 hour of the ECG.
Heart rate $>90\,\mathrm{min}^{-1}$ was derived from the
machine-reported R--R interval, with waveform peak detection
used when the interval was invalid. The WBC component was
classified from the model-estimated probability of a count
$<4$ or $>12$\,K/$\mu$L using the fixed validation-derived
threshold. The score was the sum of the four binary components,
and SIRS-positive status was defined as a score of at least 2.
The primary mortality analyses required all four components to
be available and used this binary classification.

Agreement with laboratory-defined mGPS and SIRS was assessed in all
matched ECGs for which the required laboratory tests had been
measured, without restriction to one ECG per admission or patient.
Laboratory SIRS used the same temperature, respiratory rate, and
heart rate as hybrid ECG-SIRS and replaced the model-based WBC
component with the measured count. Agreement was summarized as exact
agreement and quadratic-weighted $\kappa$; for SIRS, $\kappa$ was
computed on the 0--4 score.

Hospital admissions and patient death dates in MIMIC-IV were
linked to ECG records. For in-hospital mortality, the unit of
analysis was the hospital admission, and the first matched ECG of
each admission was retained; for hybrid ECG-SIRS, an admission was
included only if that ECG had all four components. Patients with multiple
admissions could contribute more than one admission.
In-hospital death was identified using
\texttt{hospital\_expire\_flag}.
For one-year all-cause mortality, the unit of analysis was the
patient, and the patient's first matched ECG was retained, again
requiring all four components for hybrid ECG-SIRS. Follow-up was measured from this index ECG to death
or administrative censoring at 365 days.

Associations with in-hospital mortality were estimated using
age- and sex-adjusted logistic regression. Huber--White
sandwich standard errors were clustered by patient identifier
to account for repeated admissions. Associations with one-year
mortality were estimated using age- and sex-adjusted Cox
regression, with one record per patient. Kaplan--Meier curves
and log-rank tests compared one-year survival across score
categories.


\section*{RESOURCE AVAILABILITY}
\label{resource-availability}

\begin{itemize}
\item MIMIC-IV version 3.1 hospital laboratory events and the
  MIMIC-IV-ECG Diagnostic Electrocardiogram Matched Subset version 1.0
  are available to credentialed researchers through PhysioNet under
  their respective data-use
  agreements\cite{johnson2023mimic,gow2023mimic,pollard_physionet_2026}.
  The clinical-score analyses used MIMIC-IV version 3.1 hospital
  admissions, patient vital status, and ICU charted vital signs.
  Demographics and coded diagnoses in Table~\ref{tab:population} were
  taken from the MIMIC-IV version 2.2 hospital module, and
  MIMIC-IV-ED version 2.2 stay records were used only to describe
  patient-level race and ethnicity\cite{johnson2021mimic}.
\item MC-MED is available to credentialed researchers through
  PhysioNet\cite{PhysioNet-mc-med-1.0.1,kansal2025mc} at
  \url{https://physionet.org/content/mc-med/1.0.1/} after completion
  of the required training and the PhysioNet Credentialed Health Data
  Use Agreement. The authors cannot redistribute MC-MED waveforms or
  laboratory files.
\item The supplemental information includes Figure~S1 (discrimination
  versus training volume), Figure~S2 (observed-label co-occurrence
  among the shortlisted abnormalities), Table~S1 (patient-first performance for
  all 89 abnormality directions), Table~S2 (multi-seed task-level
  performance), and Table~S3 (laboratory abnormality thresholds and
  units in MC-MED and MIMIC-IV).
\item The analysis code, task definitions, configuration files, and
  random seeds will be made publicly available on GitHub
  at \url{https://github.com/PKUDigitalHealth/ECG-LabWAS} no later than publication.
  Patient-level clinical data are not included and must be obtained
  separately through the relevant PhysioNet access procedures.
  Pretrained ECGFounder weights are available from the original
  release\cite{li2025electrocardiogram}.
\item Any additional information required to reanalyze the data
  reported in this paper is available from the corresponding author upon
  request.
\end{itemize}

\section*{ACKNOWLEDGMENTS}

This work is supported by the Beijing Nova Program (202604841257), Beijing Municipal Science and Technology Commission (Z251100000725008), Capital's Funds for Health Improvement and Research (CFH2026-1-4092), Research Project of Peking University in the State Key Laboratory of Vascular Homeostasis and Remodeling (2025-SKLVHR-YCTS-02) and the National Natural Science Foundation of China (Grant No. 82600963).

\section*{DECLARATION OF INTERESTS}

The authors declare no competing interests.

\section*{SUPPLEMENTAL INFORMATION INDEX}

\noindent\textbf{Figure S1.} Discrimination versus training volume
across the MC-MED panel.

\noindent\textbf{Figure S2.} Observed-label associations among the
33 shortlisted laboratory abnormalities.

\noindent\textbf{Table S1.} Patient-first performance across all 89
ECG-LabWAS abnormality directions.

\noindent\textbf{Table S2.} Multi-seed task-level performance.

\noindent\textbf{Table S3.} Study-defined laboratory abnormality
thresholds in MC-MED and MIMIC-IV.

\bibliography{reference}


\clearpage
\renewcommand{\thefigure}{S\arabic{figure}}
\renewcommand{\thetable}{S\arabic{table}}
\setcounter{figure}{0}
\setcounter{table}{0}

\begin{center}
{\Large\bfseries Supplemental Information}\\[0.35em]
{\large ECG-LabWAS: A Laboratory-Wide Association Study Using a
Fine-Tuned Single-Lead Electrocardiogram Foundation Model}
\end{center}

\noindent Shaded rows in Table~S1 mark the 33 directions that met the
same patient-first statistical, sample-size, and discrimination
criteria in both cohorts.

\vspace{0.4em}
\begin{center}
\includegraphics[width=0.86\linewidth]{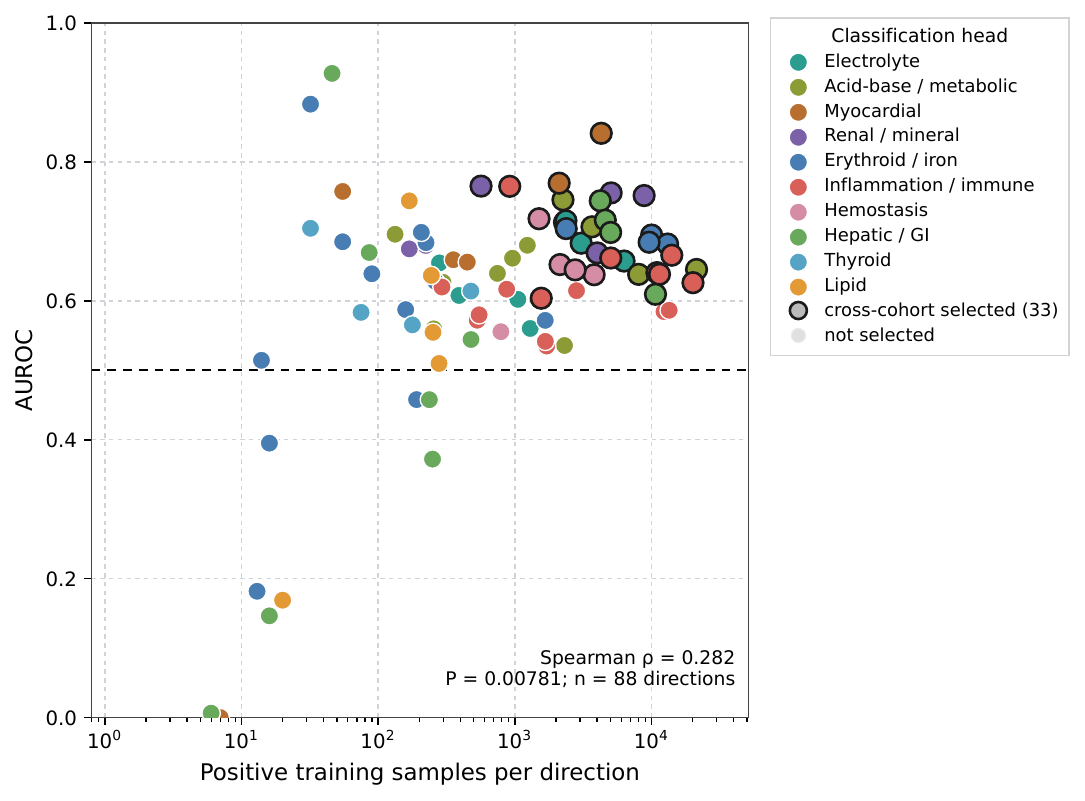}
\end{center}
\captionof{figure}{\textbf{Discrimination versus training volume
across the MC-MED panel.}
Patient-first test AUROC for the 88 evaluable abnormality directions,
plotted against the number of positive training samples on a
logarithmic scale. Colors indicate the classification
head. Outlined symbols mark the 33 directions that later met the
common criteria in both cohorts; symbols without an outline mark
directions outside the shortlist. The dashed line indicates chance discrimination. AUROC
was weakly correlated with the number of positive training
samples (Spearman $\rho=0.28$,
$P=0.008$). Extreme values, including elevated albumin
(AUROC 0.927) and low ferritin (0.883) at the top and low lactate
dehydrogenase (0.000) and low alkaline phosphatase (0.006) at the
bottom, came from directions with fewer than 50 positive training
samples and one to three positive test patients, and none of them met
the Bonferroni criterion. Low lactate had no positive test patient
and is not shown. Every shortlisted direction had at least 566 positive
training samples.}
\label{fig:s1}

\clearpage
\begin{figure}[p]
\centering
\includegraphics[width=0.82\linewidth]
{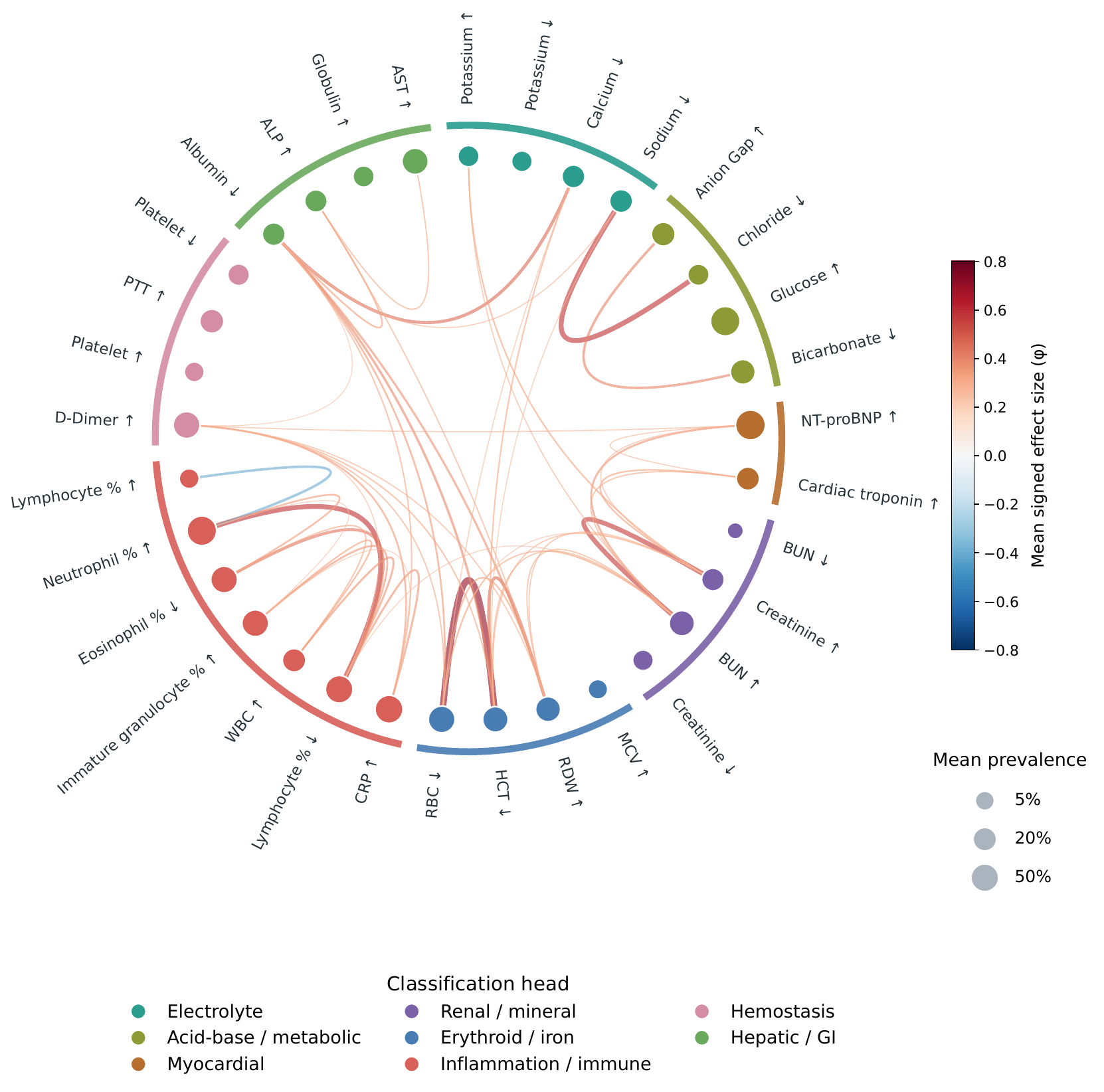}
\caption{\textbf{Associations among the 33 shortlisted
laboratory abnormalities.}
Nodes are abnormality directions grouped by classification head.
Patient-level labels indicate whether an abnormality occurred at
least once in the available laboratory history. Of 528 tested pairs,
490 were significant in both MC-MED and MIMIC-IV with the same sign
(cohort-specific Benjamini--Hochberg false discovery rate 0.05).
Edges show the 50 relationships with the largest mean absolute
$\varphi$ coefficients. Node size represents mean prevalence,
and edge color the signed mean $\varphi$ coefficient.
The strongest shared associations linked low hematocrit with low
red cell count (mean $\varphi=0.703$), low chloride with low sodium
(0.574), low lymphocyte percentage with elevated neutrophil
percentage (0.569), elevated creatinine with elevated BUN (0.552),
low hematocrit with elevated red cell distribution width (0.439),
and low albumin with low calcium (0.438). Several follow from test
definitions or measurement: hematocrit depends on red cell count
and volume, leukocyte percentages sum to 100\%, and total calcium
is partly bound to albumin. This network provides clinical context;
it does not establish ECG-specific information or novel laboratory
relationships. It summarizes observed laboratory histories, not
simultaneous measurements or model-derived associations.}
\label{fig:cooccurrence}
\end{figure}

\clearpage
\begingroup
\scriptsize
\setlength{\tabcolsep}{1.6pt}
\begin{longtable}{@{}>{\raggedright\arraybackslash}p{0.20\linewidth} l >{\raggedright\arraybackslash}p{0.17\linewidth} c r c r@{}}
\rowcolor{white}
\caption{\textbf{Performance across all 89 ECG-LabWAS abnormality
directions.}
Patient-first AUROC with stratified bootstrap 95\% confidence
intervals in MC-MED and MIMIC-IV. Each $n$ is the number of evaluable
patients, and the percentage is the positive rate among them. Rates
below 0.1\% are shown as $<$0.1\%.
Shaded rows mark the 33 directions that met the common Bonferroni
significance, class-count, total-observation, and AUROC criteria in
both cohorts. A dash indicates that no AUROC could be estimated
because the task was unmapped or a class was not evaluable; a patient
count may still be shown when the task was observed. Estimates for
the shortlisted set are descriptive because both cohorts were used to
select it. \textsuperscript{a}Troponin I in MC-MED; troponin I or troponin T in MIMIC-IV, each judged against its own threshold (see Methods).}
\label{tab:s1}\\
\toprule
Laboratory test & Dir. & Head &
\multicolumn{2}{c}{MC-MED} &
\multicolumn{2}{c}{MIMIC-IV} \\
\cmidrule(lr){4-5}\cmidrule(lr){6-7}
& & & AUROC (95\% CI) & $n$ (\%) & AUROC (95\% CI) & $n$ (\%) \\
\midrule
\endfirsthead
\caption[]{\textbf{Performance across all 89 ECG-LabWAS abnormality
directions (continued).} Shaded rows mark shortlisted directions.}\\
\toprule
Laboratory test & Dir. & Head &
\multicolumn{2}{c}{MC-MED} &
\multicolumn{2}{c}{MIMIC-IV} \\
\cmidrule(lr){4-5}\cmidrule(lr){6-7}
& & & AUROC (95\% CI) & $n$ (\%) & AUROC (95\% CI) & $n$ (\%) \\
\midrule
\endhead
\midrule
\multicolumn{7}{r}{\emph{Continued on next page}}\\
\endfoot
\bottomrule
\endlastfoot
Albumin & HIGH & Hepatic/GI & 0.927 (0.917--0.937) & 2{,}822 ($<$0.1\%) & 0.815 (0.759--0.862) & 37{,}641 (0.2\%) \\
Ferritin & LOW & Erythroid/iron & 0.883 (0.696--1.000) & 60 (5.0\%) & 0.575 (0.544--0.607) & 4{,}220 (8.6\%) \\
\rowcolor{selectedrow} NT-proBNP & HIGH & Myocardial & 0.841 (0.807--0.874) & 602 (75.4\%) & 0.838 (0.830--0.847) & 13{,}621 (84.6\%) \\
\rowcolor{selectedrow} Cardiac troponin\textsuperscript{a} & HIGH & Myocardial & 0.769 (0.733--0.803) & 1{,}516 (11.1\%) & 0.740 (0.735--0.744) & 58{,}101 (23.4\%) \\
\rowcolor{selectedrow} Blood urea nitrogen (BUN) & LOW & Renal/mineral & 0.765 (0.699--0.827) & 2{,}856 (1.4\%) & 0.740 (0.725--0.754) & 105{,}783 (1.0\%) \\
\rowcolor{selectedrow} C-reactive protein (CRP) & HIGH & Inflammation/immunity & 0.765 (0.687--0.838) & 179 (63.7\%) & 0.734 (0.715--0.752) & 3{,}150 (50.8\%) \\
Total CK & LOW & Myocardial & 0.757 (0.605--0.888) & 180 (6.1\%) & 0.562 (0.548--0.574) & 32{,}309 (5.8\%) \\
\rowcolor{selectedrow} Creatinine & HIGH & Renal/mineral & 0.755 (0.732--0.780) & 2{,}848 (13.8\%) & 0.731 (0.727--0.735) & 105{,}986 (15.8\%) \\
\rowcolor{selectedrow} Blood urea nitrogen (BUN) & HIGH & Renal/mineral & 0.752 (0.731--0.772) & 2{,}856 (24.9\%) & 0.719 (0.716--0.722) & 105{,}783 (32.3\%) \\
\rowcolor{selectedrow} Anion gap & HIGH & Acid--base/metabolic & 0.745 (0.706--0.782) & 2{,}848 (6.6\%) & 0.661 (0.657--0.665) & 102{,}396 (29.2\%) \\
\rowcolor{selectedrow} Albumin & LOW & Hepatic/GI & 0.744 (0.718--0.771) & 2{,}822 (12.3\%) & 0.727 (0.721--0.733) & 37{,}641 (20.0\%) \\
HDL cholesterol & LOW & Lipids & 0.744 (0.594--0.867) & 72 (19.4\%) & 0.600 (0.584--0.615) & 7{,}268 (20.8\%) \\
\rowcolor{selectedrow} D-dimer & HIGH & Hemostasis & 0.718 (0.666--0.771) & 371 (54.7\%) & 0.691 (0.680--0.702) & 8{,}282 (38.9\%) \\
\rowcolor{selectedrow} Alkaline phosphatase & HIGH & Hepatic/GI & 0.716 (0.690--0.743) & 2{,}821 (13.3\%) & 0.640 (0.632--0.646) & 46{,}318 (15.9\%) \\
\rowcolor{selectedrow} Potassium & HIGH & Electrolytes & 0.715 (0.676--0.753) & 2{,}851 (6.4\%) & 0.637 (0.631--0.644) & 103{,}377 (8.9\%) \\
\rowcolor{selectedrow} Potassium & LOW & Electrolytes & 0.713 (0.676--0.747) & 2{,}851 (8.6\%) & 0.710 (0.704--0.715) & 103{,}377 (7.1\%) \\
\rowcolor{selectedrow} Chloride & LOW & Acid--base/metabolic & 0.706 (0.675--0.735) & 2{,}848 (10.5\%) & 0.682 (0.676--0.688) & 103{,}625 (7.4\%) \\
Free thyroxine & LOW & Thyroid & 0.704 (0.520--0.864) & 79 (12.7\%) & 0.567 (0.520--0.612) & 2{,}960 (4.1\%) \\
\rowcolor{selectedrow} Mean corpuscular volume (MCV) & HIGH & Erythroid/iron & 0.704 (0.668--0.736) & 2{,}986 (5.9\%) & 0.654 (0.646--0.661) & 106{,}071 (4.5\%) \\
Red blood cells (RBC) & HIGH & Erythroid/iron & 0.699 (0.557--0.842) & 2{,}987 (0.5\%) & 0.631 (0.604--0.659) & 106{,}071 (0.4\%) \\
\rowcolor{selectedrow} Globulin & HIGH & Hepatic/GI & 0.698 (0.673--0.725) & 2{,}820 (15.6\%) & 0.630 (0.597--0.661) & 4{,}635 (6.9\%) \\
Whole-blood lactate & HIGH & Acid--base/metabolic & 0.696 (0.530--0.846) & 48 (45.8\%) & 0.638 (0.632--0.643) & 40{,}659 (32.3\%) \\
\rowcolor{selectedrow} Red cell distribution width (RDW) & HIGH & Erythroid/iron & 0.695 (0.672--0.715) & 2{,}985 (26.5\%) & 0.680 (0.677--0.684) & 106{,}061 (25.9\%) \\
Hematocrit (HCT) & HIGH & Erythroid/iron & 0.685 (0.375--0.989) & 2{,}993 (0.2\%) & 0.721 (0.675--0.764) & 106{,}809 (0.1\%) \\
\rowcolor{selectedrow} Hematocrit (HCT) & LOW & Erythroid/iron & 0.684 (0.662--0.705) & 2{,}993 (26.7\%) & 0.663 (0.659--0.666) & 106{,}809 (31.5\%) \\
Red cell distribution width (RDW) & LOW & Erythroid/iron & 0.684 (0.585--0.779) & 2{,}985 (0.9\%) & 0.673 (0.637--0.709) & 106{,}061 (0.2\%) \\
\rowcolor{selectedrow} Calcium & LOW & Electrolytes & 0.683 (0.652--0.717) & 2{,}849 (8.3\%) & 0.658 (0.652--0.664) & 52{,}896 (20.2\%) \\
\rowcolor{selectedrow} Red blood cells (RBC) & LOW & Erythroid/iron & 0.682 (0.663--0.701) & 2{,}987 (39.3\%) & 0.658 (0.655--0.661) & 106{,}071 (41.7\%) \\
Bicarbonate & HIGH & Acid--base/metabolic & 0.680 (0.624--0.729) & 2{,}848 (3.3\%) & 0.584 (0.578--0.589) & 102{,}758 (10.2\%) \\
Phosphorus & LOW & Renal/mineral & 0.680 (0.569--0.787) & 193 (11.4\%) & 0.586 (0.579--0.594) & 51{,}658 (11.6\%) \\
Phosphorus & HIGH & Renal/mineral & 0.675 (0.571--0.779) & 193 (10.9\%) & 0.653 (0.645--0.661) & 51{,}658 (10.9\%) \\
Plasma ammonia & LOW & Hepatic/GI & 0.669 (0.471--0.854) & 58 (12.1\%) & 0.568 (0.496--0.641) & 515 (12.4\%) \\
\rowcolor{selectedrow} Creatinine & LOW & Renal/mineral & 0.669 (0.642--0.696) & 2{,}848 (14.0\%) & 0.635 (0.626--0.643) & 105{,}986 (3.8\%) \\
\rowcolor{selectedrow} Lymphocyte \% & LOW & Inflammation/immunity & 0.666 (0.646--0.685) & 2{,}788 (47.3\%) & 0.686 (0.683--0.690) & 88{,}956 (46.6\%) \\
\rowcolor{selectedrow} White blood cells (WBC) & HIGH & Inflammation/immunity & 0.662 (0.634--0.687) & 2{,}996 (16.0\%) & 0.692 (0.687--0.696) & 106{,}132 (17.5\%) \\
Anion gap & LOW & Acid--base/metabolic & 0.661 (0.604--0.725) & 2{,}848 (2.8\%) & 0.514 (0.483--0.545) & 102{,}396 (0.3\%) \\
Total LDH & HIGH & Myocardial & 0.659 (0.511--0.792) & 65 (67.7\%) & 0.619 (0.609--0.629) & 15{,}516 (32.2\%) \\
\rowcolor{selectedrow} Sodium & LOW & Electrolytes & 0.657 (0.633--0.683) & 2{,}856 (18.2\%) & 0.631 (0.626--0.636) & 103{,}159 (11.4\%) \\
Total CK & HIGH & Myocardial & 0.656 (0.571--0.734) & 180 (35.0\%) & 0.541 (0.534--0.548) & 32{,}309 (25.7\%) \\
Sodium & HIGH & Electrolytes & 0.655 (0.543--0.758) & 2{,}856 (0.8\%) & 0.660 (0.648--0.672) & 103{,}159 (1.8\%) \\
\rowcolor{selectedrow} Platelet count (PLT) & HIGH & Hemostasis & 0.652 (0.615--0.691) & 2{,}995 (7.0\%) & 0.653 (0.646--0.660) & 106{,}251 (5.8\%) \\
\rowcolor{selectedrow} Glucose & HIGH & Acid--base/metabolic & 0.645 (0.622--0.667) & 2{,}856 (72.3\%) & 0.645 (0.642--0.649) & 104{,}859 (66.9\%) \\
\rowcolor{selectedrow} Partial thromboplastin time & HIGH & Hemostasis & 0.645 (0.609--0.680) & 1{,}238 (23.6\%) & 0.633 (0.628--0.638) & 69{,}229 (17.9\%) \\
\rowcolor{selectedrow} Immature granulocyte \% & HIGH & Inflammation/immunity & 0.641 (0.618--0.661) & 2{,}786 (36.9\%) & 0.635 (0.629--0.642) & 30{,}951 (42.8\%) \\
Serum osmolality & HIGH & Acid--base/metabolic & 0.639 (0.572--0.709) & 247 (40.5\%) & 0.615 (0.593--0.638) & 2{,}353 (41.5\%) \\
TIBC & LOW & Erythroid/iron & 0.639 (0.425--0.829) & 33 (36.4\%) & 0.611 (0.591--0.630) & 3{,}905 (24.7\%) \\
\rowcolor{selectedrow} Eosinophil \% & LOW & Inflammation/immunity & 0.638 (0.619--0.659) & 2{,}788 (40.5\%) & 0.651 (0.647--0.655) & 88{,}947 (36.9\%) \\
\rowcolor{selectedrow} Bicarbonate & LOW & Acid--base/metabolic & 0.638 (0.616--0.663) & 2{,}848 (25.6\%) & 0.661 (0.657--0.666) & 102{,}758 (21.4\%) \\
\rowcolor{selectedrow} Platelet count (PLT) & LOW & Hemostasis & 0.638 (0.609--0.668) & 2{,}995 (10.3\%) & 0.648 (0.643--0.653) & 106{,}251 (10.7\%) \\
Triglycerides & HIGH & Lipids & 0.636 (0.476--0.778) & 73 (26.0\%) & 0.562 (0.549--0.575) & 7{,}319 (34.3\%) \\
Ferritin & HIGH & Erythroid/iron & 0.628 (0.491--0.756) & 60 (50.0\%) & 0.628 (0.611--0.645) & 4{,}220 (28.1\%) \\
Serum osmolality & LOW & Acid--base/metabolic & 0.627 (0.538--0.710) & 247 (17.8\%) & 0.654 (0.629--0.679) & 2{,}353 (27.1\%) \\
\rowcolor{selectedrow} Neutrophil \% & HIGH & Inflammation/immunity & 0.626 (0.603--0.647) & 2{,}788 (70.1\%) & 0.639 (0.635--0.643) & 88{,}948 (74.6\%) \\
Monocyte \% & LOW & Inflammation/immunity & 0.620 (0.488--0.742) & 2{,}788 (0.9\%) & 0.645 (0.634--0.656) & 88{,}948 (3.1\%) \\
Procalcitonin & HIGH & Inflammation/immunity & 0.617 (0.479--0.762) & 143 (86.0\%) & --- & --- \\
Eosinophil \% & HIGH & Inflammation/immunity & 0.615 (0.579--0.647) & 2{,}788 (10.6\%) & 0.599 (0.594--0.605) & 88{,}947 (10.1\%) \\
TSH & HIGH & Thyroid & 0.614 (0.540--0.687) & 379 (19.8\%) & 0.586 (0.574--0.599) & 14{,}241 (15.5\%) \\
\rowcolor{selectedrow} AST (SGOT) & HIGH & Hepatic/GI & 0.609 (0.589--0.631) & 2{,}821 (37.2\%) & 0.620 (0.615--0.625) & 47{,}905 (36.9\%) \\
Magnesium & LOW & Electrolytes & 0.608 (0.510--0.697) & 567 (6.0\%) & 0.639 (0.631--0.647) & 54{,}021 (9.9\%) \\
\rowcolor{selectedrow} Lymphocyte \% & HIGH & Inflammation/immunity & 0.604 (0.561--0.644) & 2{,}788 (6.5\%) & 0.640 (0.634--0.647) & 88{,}956 (7.6\%) \\
Calcium & HIGH & Electrolytes & 0.602 (0.540--0.663) & 2{,}849 (2.8\%) & 0.609 (0.597--0.621) & 52{,}896 (4.1\%) \\
Transferrin saturation & LOW & Erythroid/iron & 0.587 (0.381--0.799) & 30 (70.0\%) & --- & --- \\
Basophil \% & LOW & Inflammation/immunity & 0.586 (0.566--0.608) & 2{,}788 (48.9\%) & 0.600 (0.596--0.604) & 88{,}948 (48.8\%) \\
Monocyte \% & HIGH & Inflammation/immunity & 0.584 (0.563--0.605) & 2{,}788 (44.0\%) & 0.531 (0.525--0.536) & 88{,}948 (17.1\%) \\
Free thyroxine & HIGH & Thyroid & 0.583 (0.425--0.758) & 79 (8.9\%) & 0.673 (0.629--0.711) & 2{,}960 (6.7\%) \\
Automated ESR & HIGH & Inflammation/immunity & 0.580 (0.478--0.679) & 123 (58.5\%) & 0.654 (0.629--0.682) & 1{,}691 (57.2\%) \\
Neutrophil \% & LOW & Inflammation/immunity & 0.572 (0.505--0.637) & 2{,}788 (2.0\%) & 0.556 (0.542--0.568) & 88{,}948 (2.3\%) \\
Mean corpuscular volume (MCV) & LOW & Erythroid/iron & 0.572 (0.517--0.621) & 2{,}986 (4.6\%) & 0.583 (0.575--0.591) & 106{,}071 (5.1\%) \\
TSH & LOW & Thyroid & 0.565 (0.440--0.687) & 379 (4.5\%) & 0.571 (0.553--0.591) & 14{,}241 (6.3\%) \\
Magnesium & HIGH & Electrolytes & 0.560 (0.502--0.612) & 567 (26.3\%) & 0.570 (0.563--0.577) & 54{,}021 (17.7\%) \\
Glucose & LOW & Acid--base/metabolic & 0.560 (0.447--0.692) & 2{,}856 (0.7\%) & 0.556 (0.541--0.572) & 104{,}859 (1.3\%) \\
Partial thromboplastin time & LOW & Hemostasis & 0.555 (0.503--0.609) & 1{,}238 (8.2\%) & 0.557 (0.551--0.562) & 69{,}229 (17.1\%) \\
Calculated non-HDL cholesterol & HIGH & Lipids & 0.555 (0.410--0.697) & 72 (29.2\%) & --- & --- \\
Globulin & LOW & Hepatic/GI & 0.544 (0.465--0.624) & 2{,}820 (1.7\%) & 0.501 (0.465--0.536) & 4{,}635 (5.9\%) \\
Basophil \% & HIGH & Inflammation/immunity & 0.542 (0.498--0.583) & 2{,}788 (5.6\%) & 0.552 (0.545--0.558) & 88{,}948 (8.3\%) \\
Chloride & HIGH & Acid--base/metabolic & 0.536 (0.493--0.577) & 2{,}848 (6.3\%) & 0.474 (0.468--0.479) & 103{,}625 (12.6\%) \\
White blood cells (WBC) & LOW & Inflammation/immunity & 0.535 (0.481--0.588) & 2{,}996 (5.1\%) & 0.550 (0.541--0.558) & 106{,}132 (3.7\%) \\
TIBC & HIGH & Erythroid/iron & 0.514 (0.250--0.793) & 33 (15.2\%) & 0.526 (0.494--0.560) & 3{,}905 (7.9\%) \\
Direct LDL cholesterol & HIGH & Lipids & 0.510 (0.377--0.640) & 72 (34.7\%) & 0.591 (0.568--0.617) & 1{,}983 (50.4\%) \\
Total iron & LOW & Erythroid/iron & 0.458 (0.236--0.689) & 34 (73.5\%) & 0.593 (0.576--0.611) & 4{,}147 (63.4\%) \\
Lipase & HIGH & Hepatic/GI & 0.458 (0.361--0.561) & 698 (3.4\%) & 0.548 (0.532--0.565) & 29{,}805 (4.5\%) \\
Transferrin saturation & HIGH & Erythroid/iron & 0.395 (0.074--0.815) & 30 (10.0\%) & --- & --- \\
Plasma ammonia & HIGH & Hepatic/GI & 0.372 (0.225--0.533) & 58 (25.9\%) & 0.573 (0.520--0.622) & 515 (49.5\%) \\
Total iron & HIGH & Erythroid/iron & 0.182 (0.061--0.333) & 34 (2.9\%) & 0.554 (0.508--0.604) & 4{,}147 (3.3\%) \\
HDL cholesterol & HIGH & Lipids & 0.169 (0.085--0.268) & 72 (1.4\%) & 0.560 (0.516--0.602) & 7{,}268 (2.2\%) \\
AST (SGOT) & LOW & Hepatic/GI & 0.146 (0.134--0.160) & 2{,}821 ($<$0.1\%) & 0.525 (0.439--0.605) & 47{,}905 (0.1\%) \\
Alkaline phosphatase & LOW & Hepatic/GI & 0.006 (0.004--0.010) & 2{,}821 ($<$0.1\%) & 0.517 (0.412--0.617) & 46{,}318 (0.1\%) \\
Total LDH & LOW & Myocardial & 0.000 (0.000--0.000) & 65 (1.5\%) & 0.553 (0.533--0.575) & 15{,}516 (4.0\%) \\
Whole-blood lactate & LOW & Acid--base/metabolic & --- & 48 (0.0\%) & 0.512 (0.387--0.622) & 40{,}659 (0.1\%) \\
\end{longtable}
\endgroup

\noindent{\scriptsize AST, aspartate aminotransferase; CK, creatine
kinase; ESR, erythrocyte sedimentation rate; GI, gastrointestinal;
HDL, high-density lipoprotein; LDH, lactate dehydrogenase; LDL,
low-density lipoprotein; SGOT, serum glutamic-oxaloacetic
transaminase; TIBC, total iron-binding capacity; TSH,
thyroid-stimulating hormone.}

\clearpage
\begin{center}
\captionof{table}{\textbf{Multi-seed task-level performance.}
MC-MED fine-tuning was repeated on the same patient partition with
seeds 42, 123, and 456. Each validation-selected checkpoint was evaluated on the same 47 mapped MIMIC-IV tasks (300{,}035 ECGs
from 115{,}921 patients). Values are sample-level task-macro
abnormal-class AUROCs, averaged over 50 MC-MED tasks and 47 mapped
MIMIC-IV tasks, respectively, without patient-first reduction.
Best epochs are numbered from 1. The final row reports the mean and
sample standard deviation across three training seeds; SD is not a
95\% confidence interval.
Seed~42 was used for the primary direction-level,
matching-window, coupling, and clinical-score analyses. The
co-occurrence analysis used observed labels for the shortlist defined
by the seed-42 model.}
\label{tab:s2}
\vspace{0.5em}
\begin{tabular}{@{}lccc@{}}
\toprule
Seed & Best epoch & MC-MED test AUROC & MIMIC-IV AUROC \\
\midrule
42 & 3 & 0.617 & 0.625 \\
123 & 2 & 0.611 & 0.627 \\
456 & 3 & 0.618 & 0.623 \\
\midrule
Mean $\pm$ SD & --- & $0.615\pm 0.004$ & $0.625\pm 0.002$ \\
\bottomrule
\end{tabular}
\end{center}

\clearpage
\begingroup
\footnotesize
\setlength{\tabcolsep}{3pt}
\renewcommand{\arraystretch}{1.15}
\begin{longtable}{@{}>{\raggedright\arraybackslash}p{0.27\linewidth} >{\raggedright\arraybackslash}p{0.12\linewidth} >{\centering\arraybackslash}p{0.105\linewidth} >{\centering\arraybackslash}p{0.105\linewidth} >{\centering\arraybackslash}p{0.105\linewidth} >{\centering\arraybackslash}p{0.185\linewidth}@{}}
\caption{\textbf{Study-defined laboratory abnormality thresholds in MC-MED and MIMIC-IV.}
Thresholds define the 50 model tasks, not universal clinical reference intervals.
Units are the normalized classification units. LOW and HIGH use strict
inequalities; equality is NORMAL. A dash indicates that no LOW class
was defined; NM denotes an unmapped MIMIC-IV task.}
\label{tab:thresholds}\\
\toprule
\rowcolor{white}
Laboratory task & Unit & \multicolumn{2}{c}{MC-MED} & \multicolumn{2}{c}{MIMIC-IV} \\
\cmidrule(lr){3-4}\cmidrule(l){5-6}
\rowcolor{white}
 & & LOW & HIGH & LOW & HIGH \\
\midrule
\endfirsthead
\caption[]{\textbf{Study-defined laboratory abnormality thresholds (continued).}}\\
\toprule
\rowcolor{white}
Laboratory task & Unit & \multicolumn{2}{c}{MC-MED} & \multicolumn{2}{c}{MIMIC-IV} \\
\cmidrule(lr){3-4}\cmidrule(l){5-6}
\rowcolor{white}
 & & LOW & HIGH & LOW & HIGH \\
\midrule
\endhead
\bottomrule
\endfoot
\rowcolor{gray!10}
\multicolumn{6}{@{}l}{\textbf{H1: Electrolytes}} \\*
Potassium & mmol/L & $<3.5$ & $>5$ & $<3.5$ & $>5$ \\
Calcium & mg/dL & $<8.5$ & $>10.2$ & $<8.5$ & $>10.2$ \\
Magnesium & mg/dL & $<1.7$ & $>2.2$ & $<1.7$ & $>2.2$ \\
Sodium & mmol/L & $<135$ & $>145$ & $<135$ & $>145$ \\
\rowcolor{gray!10}
\multicolumn{6}{@{}l}{\textbf{H2: Acid--base/metabolic}} \\*
Bicarbonate & mmol/L & $<23$ & $>29$ & $<23$ & $>29$ \\
Chloride & mmol/L & $<96$ & $>106$ & $<96$ & $>106$ \\
Whole-blood lactate & mmol/L & $<0.5$ & $>2.2$ & $<0.5$ & $>2.2$ \\
Glucose & mg/dL & $<70$ & $>99$ & $<70$ & $>99$ \\
Serum osmolality & mOsm/kg & $<275$ & $>295$ & $<275$ & $>295$ \\
Anion gap & mmol/L & $<8$ & $>16$ & $<8$ & $>16$ \\
\rowcolor{gray!10}
\multicolumn{6}{@{}l}{\textbf{H3: Myocardial}} \\*
NT-proBNP & pg/mL & --- & $>125$ & --- & $>125$ \\
Cardiac troponin\textsuperscript{a} & ng/mL & --- & $>0.04$ & --- & \makecell[t]{$>0.04$ (TnI)\\$>0.01$ (TnT)} \\
Total creatine kinase & U/L & $<30$ & $>200$ & $<30$ & $>200$ \\
Total LDH & U/L & $<140$ & $>280$ & $<140$ & $>280$ \\
\rowcolor{gray!10}
\multicolumn{6}{@{}l}{\textbf{H4: Renal/mineral}} \\*
Creatinine & mg/dL & $<0.6$ & $>1.3$ & $<0.6$ & $>1.3$ \\
BUN & mg/dL & $<6$ & $>20$ & $<6$ & $>20$ \\
Phosphorus & mg/dL & $<2.5$ & $>4.5$ & $<2.5$ & $>4.5$ \\
\rowcolor{gray!10}
\multicolumn{6}{@{}l}{\textbf{H5: Erythroid/iron}} \\*
Red blood cell count & $10^{6}/\mu$L & $<4.2$ & $>6.1$ & $<4.2$ & $>6.1$ \\
Hematocrit & \% & $<36$ & $>55$ & $<36$ & $>55$ \\
Mean corpuscular volume & fL & $<80$ & $>100$ & $<80$ & $>100$ \\
Red cell distribution width & \% & $<11.5$ & $>14.5$ & $<11.5$ & $>14.5$ \\
Total iron & $\mu$g/dL & $<60$ & $>170$ & $<60$ & $>170$ \\
Total iron-binding capacity & $\mu$g/dL & $<240$ & $>450$ & $<240$ & $>450$ \\
Ferritin & ng/mL & $<12$ & $>300$ & $<12$ & $>300$ \\
Transferrin saturation & \% & $<20$ & $>50$ & NM & NM \\
\rowcolor{gray!10}
\multicolumn{6}{@{}l}{\textbf{H6: Inflammation/immunity}} \\*
White blood cell count & $10^{3}/\mu$L & $<4$ & $>12$ & $<4$ & $>12$ \\
Neutrophil \% & \% & $<40$ & $>60$ & $<40$ & $>60$ \\
Lymphocyte \% & \% & $<20$ & $>40$ & $<20$ & $>40$ \\
Monocyte \% & \% & $<2$ & $>8$ & $<2$ & $>8$ \\
Eosinophil \% & \% & $<1$ & $>4$ & $<1$ & $>4$ \\
Basophil \% & \% & $<0.5$ & $>1$ & $<0.5$ & $>1$ \\
Immature granulocyte \% & \% & --- & $>0.4$ & --- & $>0.4$ \\
C-reactive protein & mg/dL & --- & $>1$ & --- & $>1$ \\
Automated ESR & mm/h & --- & $>20$ & --- & $>20$ \\
Procalcitonin & ng/mL & --- & $>0.05$ & NM & NM \\
\rowcolor{gray!10}
\multicolumn{6}{@{}l}{\textbf{H7: Hemostasis}} \\*
Platelet count & $10^{3}/\mu$L & $<150$ & $>400$ & $<150$ & $>400$ \\
Partial thromboplastin time & s & $<25$ & $>35$ & $<25$ & $>35$ \\
D-dimer & $\mu$g/mL FEU & --- & $>0.5$ & --- & $>0.5$ \\
\rowcolor{gray!10}
\multicolumn{6}{@{}l}{\textbf{H8: Hepatic/GI}} \\*
Albumin & g/dL & $<3.5$ & $>5.4$ & $<3.5$ & $>5.4$ \\
Globulin & g/dL & $<2$ & $>3.5$ & $<2$ & $>3.5$ \\
AST & U/L & $<8$ & $>33$ & $<8$ & $>33$ \\
Alkaline phosphatase & U/L & $<20$ & $>130$ & $<20$ & $>130$ \\
Plasma ammonia & $\mu$mol/L & $<15$ & $>45$ & $<15$ & $>45$ \\
Lipase & U/L & --- & $>160$ & --- & $>160$ \\
\rowcolor{gray!10}
\multicolumn{6}{@{}l}{\textbf{H9: Thyroid}} \\*
TSH & $\mu$IU/mL & $<0.4$ & $>4$ & $<0.4$ & $>4$ \\
Free thyroxine & ng/dL & $<0.8$ & $>1.8$ & $<0.8$ & $>1.8$ \\
\rowcolor{gray!10}
\multicolumn{6}{@{}l}{\textbf{H10: Lipids}} \\*
HDL cholesterol & mg/dL & $<40$ & $>100$ & $<40$ & $>100$ \\
Direct LDL cholesterol & mg/dL & --- & $>100$ & --- & $>100$ \\
Calculated non-HDL cholesterol & mg/dL & --- & $>130$ & NM & NM \\
Triglycerides & mg/dL & --- & $>150$ & --- & $>150$ \\
\end{longtable}
\endgroup

\noindent{\footnotesize\textsuperscript{a}Cardiac troponin retains the model key
\texttt{TROPONIN I}. MC-MED training uses TnI $>0.04$\,ng/mL.
MIMIC-IV combines TnI (itemid 52642; itemid 51002 had no records)
at $>0.04$\,ng/mL and TnT (itemid 51003) at $>0.01$\,ng/mL.
If both are available for an ECG, the result nearest in time is used.
For classification only, TnT values are multiplied by four so that one
threshold applies; reported values and units are unchanged. This aligns
thresholds, not assay values.}

\noindent{\footnotesize NORMAL is the closed interval between the listed
bounds for ternary tasks and $x\leq U$ for binary tasks, where $U$ is
the HIGH threshold. Zero-valued lower-bound placeholders for binary
tasks do not define a LOW class. Transferrin saturation, calculated
non-HDL cholesterol, and procalcitonin were not mapped in MIMIC-IV
and remained masked, not normal. These fixed task-level thresholds
were not made age- or sex-specific.}

\noindent{\footnotesize Unit handling: CRP in MIMIC-IV is converted
from mg/L to mg/dL by dividing by 10 when required
($1$\,mg/dL $=10$\,mg/L). D-dimer is expressed in $\mu$g/mL FEU
($0.5$\,$\mu$g/mL FEU $=500$\,ng/mL FEU); MC-MED ELISA and
MIMIC-IV results in ng/mL FEU are divided by 1,000.
TSH in mIU/L is numerically equivalent to $\mu$IU/mL.
The blood-count units $10^3/\mu$L and $10^6/\mu$L correspond to
the pipeline units K/$\mu$L and MIL/$\mu$L, respectively.
In MIMIC-IV, a censored text result is labeled only if its entire
possible interval falls in one class; ambiguous intervals remain missing.}

\end{document}